\documentclass[tradiabstract]{aa}  
\usepackage{graphicx}
\usepackage{txfonts}
\usepackage{hyperref}
\usepackage{array}
\usepackage{multirow}
\usepackage{amsbsy} 
\newcolumntype{P}[1]{>{\centering\arraybackslash}p{#1}}
\newcolumntype{M}[1]{>{\centering\arraybackslash}m{#1}}
\newcolumntype{N}{@{}m{0pt}@{}}

\graphicspath{{figures/}}
\hypersetup{ bookmarks=true, unicode=true, pdftoolbar=true, pdfmenubar=true, pdffitwindow=true, pdfstartview={FitH}, pdfauthor={SJ}, colorlinks=true, linkcolor=blue,citecolor=blue, urlcolor=black, breaklinks}

\begin{document}

\title{Constraining cosmology with the cosmic microwave and infrared backgrounds correlation}

\author{A. Maniyar
	\inst{1}
	\and
	G. Lagache
	\inst{1}
	\and
	M. B\'ethermin
	\inst{1}
	\and
	S. Ili\'c
	\inst{2} }

\institute{ $^1$Aix Marseille Univ, CNRS, CNES, LAM, Marseille, France\\
	$^2$ CEICO, Institute of Physics of the Czech Academy of Sciences, Na Slovance 2, Praha 8 Czech Republic\\
	\email{\href{mailto:abhishek.maniyar@lam.fr}{\textrm{abhishek.maniyar@lam.fr}}}
	\label{inst1}
}

\date{Received July 03, 2018; accepted xx;}

\abstract
{We explore the use of the cosmic infrared background (CIB) as a tracer of the large scale structures for cross-correlating with the cosmic microwave background (CMB) and exploit the integrated Sachs Wolfe (ISW) effect. We use the improved linear CIB model of \cite{Maniyar_2018} and derive the theoretical \mbox{CIBxISW} cross-correlation for different Planck HFI frequencies (217, 353, 545 and 857\,GHz) and IRAS (3000\,GHz). As is expected, we predict a positive cross-correlation between the CIB and the CMB whose amplitude decreases rapidly at small scales. We perform a signal-to-noise ratio (SNR) analysis on the predicted cross-correlation. In the ideal case of the cross-correlation obtained over 70\% (40\%) of the sky with no residual contaminants (e.g. galactic dust) in maps, the SNR ranges from  4.2 - 5.6 (3.2 - 4.3) with the highest SNR coming from 857\,GHz. A Fisher matrix analysis shows that an ISW signal detected with such high SNR on the 40\% sky can improve the constraints on the cosmological parameters considerably; constraints on the equation of state of the dark energy especially are improved by 80\%. We then perform a more realistic analysis considering the effect of residual galactic dust contamination in CIB maps. We calculate the dust power spectra for different frequencies and sky fractions, which dominate over the CIB power spectra at the lower multipoles we are interested in. Considering a conservative 10\% residual level of galactic dust in the CIB power spectra, we observe that the SNR drops drastically making the ISW detection very challenging. To check the capability of current maps to detect the ISW effect via this method, we measure the cross-correlation of the CIB and the CMB Planck maps on the so-called GASS field which covers an area of $\sim 11\%$ in the southern hemisphere. We find that with such a small sky fraction and  the dust residuals present in the CIB maps, we do not detect any ISW signal and the measured cross-correlation is consistent with zero. In order not to degrade the SNR for the ISW measurement by more than 10\% on the 40\% sky, we find that the dust needs to be cleaned up to the 0.01\% level on the power spectrum.}


\authorrunning{Maniyar et al.}
\titlerunning{ISW detection using CMB and CIB correlation}
\maketitle


\section{Introduction}\label{sec:1}

At the end of the last century, \cite{Rees_1998} and \cite{Perlmutter_1999} discovered that the expansion of the Universe is accelerating. Since then, a great wealth of theories have been put forward to explain this acceleration. The leading theory attributes it to  "Dark Energy", an unknown component constituting around 70\% of the total energy budget of the Universe according to our latest estimates \citep{Planck_cosmo_2016}. The $\Lambda$CDM model, also called the concordance model, is the "most favoured" model by the data in the context of Bayesian model comparison \citep{Planck_cosmo_2016}. Currently, a vast amount of effort is dedicated to constraining better the parameters of our cosmological model. Various cosmological probes of the distance scales (all the standard rulers and candles i.e. baryonic acoustic oscillations \citep{Eisenstein_1998}, supernovae Ia \citep{Betoule_2014}, cepheids \citep{Riess_2016}) and probes of the growth of structures (weak lensing \citep{Lewis_2006}, galaxy clustering \citep{Tinker_2012}, redshift space distortions \citep{Scoccimarro_2004}) are being used to study and test cosmological models such as $\Lambda$CDM and possible deviations from it.

Amongst these, the integrated Sachs-Wolfe (ISW) effect is another potential powerful probe of dark energy. It is the low-redshift ``integrated'' counterpart of the original Sachs-Wolfe effect \citep{Sachs_1967} which occurs at the surface of the last scattering and leaves its imprints on the cosmic microwave background (CMB) anisotropies. This effect is caused by the gravitational redshifting of the photons while coming out of the potential wells at the last scattering surface. The ISW effect, on the other hand occurs in a universe not dominated by matter. As dark energy starts dominating the dynamics of the Universe at late times, it causes the gravitational potential wells and hills on large scales to decay. As a result, the CMB photons travelling across them have a net gain or loss of energy for potential wells and hills, respectively. The ISW effect is the dominant contribution to the CMB power spectrum at large angular scales and has a very small amplitude. Unfortunately, at these scales, the statistical noise due to cosmic variance in the CMB power spectrum is of the same order of magnitude as the signal, making the detection of the ISW effect in the CMB alone extremely challenging.

However, we know that the matter in the Universe shapes the underlying gravitational potentials responsible for the ISW effect. Therefore, a non-zero spatial correlation between the CMB anisotropies and tracers of the matter distribution is expected, and can be exploited to extract the ISW signal \citep[e.g.][]{Boughn_1998, Boughn_2002}. Galaxy surveys (which are proxies for large scale structures) are an obvious choice for this purpose. Attempts to detect the ISW signal have been made with various surveys, such as WISE \citep{Shajib_2016,Ferraro_2015}, SDSS \citep{Cabre_2006,Giannantonio_2006,Carlos_2014}, NVSS (\mbox{\citealp{Raccanelli_2008}}, \citealp{Pietrobon_2006}) and 2MASS \citep{Francis_2010}.
Joint analyses combining different data sets have also been performed \citep{Giannantonio_2008,Planck_isw_2016,Stolzner_2017}. Alternative approaches have also been explored in the literature such as stacking CMB patches at the location of superstructures expected to yield the strongest ISW signal \citep{Ilic_2013} and exploiting the cross-correlation between the ISW and thermal Sunyaev-Zel'dovich effects -- both present in the CMB and taking place in the same gravitational wells \citep{Taburet_2011,Creque_2016}. The significance of the ISW signals detected in all those studies ranges from 1$\sigma$ to 5$\sigma$, with the main limiting factor being the lack of sky coverage and/or the redshift depth of the surveys used. Those are both crucial factors in the detection of the ISW effect, as its main contribution comes from the largest structures and thus requires the largest survey volume possible.
 
In this paper, we consider the cosmic infrared background (CIB) as a tracer of the large scale structures to probe the ISW effect. The CIB is the weighted integral of the dust heated by the young UV-bright stars within the galaxies through cosmic time. It was first detected by \cite{Puget_1996}. As it traces the star formation history of the Universe, it spans a large redshift range of $0<z<6$. Therefore, the CIB is an exceptional tool to trace the overall distribution of the galaxies at these high redshifts \citep{Knox_2001}.  \cite{Lagache_2000} and \cite{Matsuhara_2000} first detected and discussed the anisotropies in the CIB due to unresolved extra-galactic sources. Correlated anisotropies in the CIB were discovered by Spitzer \citep{Lagache_2007} and accurately measured by Planck and Herschel \citep{Planck_cib_2014, Viero_2013}. These anisotropies trace the large scale distribution of the galaxy density field and hence, the underlying distribution of the dark matter haloes which host these galaxies, up to a bias factor. Therefore, the CIB can be used as a tracer of the large scale structures \citep[e.g.][]{Hanson_2013}. 

As the CIB traces the evolution of the large scale structures over a large range of redshifts, it is a prime choice of tracer for detecting the ISW effect. \cite{Ilic_2011} showed that the \mbox{CIBxISW} cross-correlation can provide us with the highest signal-to-noise ratio for a single tracer for the detection of the ISW signal.  Since then, our understanding and modelling of the CIB anisotropies has improved \citep[e.g.][]{Bethermin_2013,Viero_2013_2,Wu_2018}, and we have gained new insights on their contaminants/foregrounds and thus the intricacies of extracting the CIB signal \citep{Planck_cib_2014}. We thus revisit the prediction for the cross-correlation between the CIB and the CMB. \cite{Maniyar_2018} considered simultaneously the CIB auto- and cross-power spectra as well as the cross-correlation of the \mbox{CIBxCMB} lensing measurements to constrain an updated model for the CIB anisotropies. Here we use this model and predict the expected signal-to-noise ratio (SNR) for the ISW effect obtained through the \mbox{CIBxCMB} cross-correlation. We then perform a Fisher matrix analysis to predict the constraints on the cosmological parameters coming from a combination of the \mbox{CIBxISW} information with CMB, CIB, and \mbox{CIBxCMB} lensing data. Unfortunately, the current CIB maps are limited by the dust contamination, especially at lower angular multipoles where the ISW signal is expected to be the strongest. Therefore, we perform our analysis in two scenarios; one with the ideal situation with dust-free CIB spectra extracted over the full or partial sky, and the second case with the CIB maps being contaminated by residual galactic dust emission.

This paper is structured as follows. We begin in Sect.~\ref{sec:2} by introducing the modelling of the various cosmological probes considered in our analysis: our model for the CIB power spectrum, the \mbox{CIBxCMB} lensing correlation, and the \mbox{CIBxISW} correlation. In Sect.~\ref{sec:3}, we provide the predictions for the ISW signal and its SNR for different cases (maps with and without residual dust contamination). In Sect.~\ref{sec:4} we explain the Fisher matrix analysis used to constrain the cosmological parameters and provide the results for the $\Lambda$CDM and $w$CDM models. Sect.~\ref{sec:5} details the measurement of the cross-correlation using the available CIB and CMB Planck maps over a small portion of the sky and the corresponding results. Conclusions are provided in Sect.~\ref{sec:6}.
For all our analysis, we are using Planck 2015 best fit cosmology \citep{Planck_cosmo_2016} with adiabatic scalar perturbations and nearly scale invariant initial power spectrum.


\section{Cosmological observables and their modelling}\label{sec:2} 


\subsection{CIB power spectrum}\label{ssec:2.1}

We aim at exploiting the cross-correlation of the CIB with the CMB through the ISW effect, which is mostly significant on large angular scales. At these scales, the clustering signal is dominated by the correlation between the dark matter halos and it is safe to ignore the non-linear effects. Thus, a linear model for calculating the CIB anisotropies is good enough for our study. The CIB power spectrum is defined as:
\begin{equation}\label{eq:1}
	C_l^{\nu \times \nu'} \times \delta _{ll'}\delta_{mm'}= \left \langle \delta I_{lm}^{\nu} \delta I_{l'm'}^{\nu'} \right \rangle \,.
\end{equation}
The CIB anisotropies for a given frequency $\nu$ at redshift $z$ and in a given direction \textbf{\^{n}} are:
\begin{equation}\label{eq:2}
\delta I_\mathrm{CIB}(\hat{\mathbf{n}},\nu,z) = \int dz \dfrac{d\chi}{dz}\, a(z)\, \delta j(\hat{\mathbf{n}},\nu,z)\,.
\end{equation}
where $I_\mathrm{CIB}$ is the CIB mean level, $\chi(z)$ is the comoving distance to redshift $z$ and $a = 1/(1+z)$ is the scale factor, and $j$ is the comoving emissivity of the CIB galaxies.  

We use the CIB model presented in \cite{Maniyar_2018}. In order to fit for the model parameters, only CIB data in the range 145 < $\ell$ < 592 are used, where the linear model is a good approximation and the Limber approximation is accurate enough and simplifies the calculation. The linear CIB anisotropy power spectrum with the Limber approximation is then given as:
\begin{equation}\label{eq:clcib}
C_l^{\nu \times \nu'} = \int \dfrac{dz}{\chi^2} \dfrac{d\chi}{dz}a^2 b_{eff}^2 \bar{j}(\nu,z) \bar{j}(\nu',z)P_{lin}(k = l/\chi,z)\,.
\end{equation}
where $P_{lin}(k,z)$ is the linear theory dark matter power spectrum which has been generated using the publicly available Boltzmann code {\tt CAMB} \citep{Lewis:2002ah, camb_notes} at the required redshifts. $b_{eff}$ is the effective bias factor for the CIB galaxies at a given redshift. It is the mean bias of dark matter halos hosting dusty galaxies contributing to the CIB at a given redshift weighted by their contribution to the emissivities (\cite{Planck_cib_2014} and \citet{Maniyar_2018}). It implies that more massive halos are more clustered and have host galaxies emitting more far-infrared emission.  We used a parametric form for the evolution of the effective bias with the redshift:
\begin{equation} \label{eq:beff}
b_{eff} (z)= b_0 + b_1z + b_2z^2 \,.
\end{equation}
It can be noted that the bias depends only on the redshift and is scale independent. This is a good approximation at the scales where we are modelling the CIB. \\
$\bar{j}$ is the comoving emissivity (in Jy\,Mpc$^{-1}$), which is derived using the star formation rate density $\rho_{SFR}$ (M$_{\odot}$\,yr$^{-1}$\,Mpc$^{-3}$) following
\begin{equation}\label{eq:jnu}
\bar{j}(\nu,z) = \dfrac{\rho_{\textrm{SFR}}(z)(1+z)S_{\nu,eff}(z)\chi^2(z)}{K}
\end{equation}
where $K$ is the Kennicutt constant \citep{Kennicutt_1998}. We used a Kennicutt constant corresponding to a Chabrier initial mass function ($ \textrm{SFR}/L_{IR} = 1.0 \times 10^{-10} \textrm{M}_{\odot}\textrm{yr}^{-1}$). $S_{\nu,eff}(z)$ (in Jy L$_{\odot}^{-1}$) is the average spectral energy distribution (SED) of all the galaxies at a given redshift weighted by their contribution to the CIB. As mentioned in \cite{Maniyar_2018}, they have been computed using the method presented in \citet{Bethermin_2013}, but assuming the new updated SEDs calibrated with the \textit{Herschel} data and presented in \citet{Bethermin_2015} and \citet{Bethermin_2017}. The parametric form of the cosmic star formation rate density proposed by \cite{Madau_2014} is used to describe $\rho_{SFR}$ and is given by:
\begin{equation}
\rho_{\textrm{SFR}}(z) = \alpha\dfrac{(1+z)^{\beta}}{1+ {[ (1+z)/{\gamma} ]}^{\delta}} \textrm{M}_{\odot} \textrm{year}^{-1} \textrm{Mpc}^{-3}\, ,
\end{equation}
where $\alpha$, $\beta$, $\gamma$ and $\delta$ are parameters of the CIB model.


\subsection{CIBxCMB lensing}\label{ssec:2.2}

As mentioned earlier, the \mbox{CIBxCMB} lensing cross-correlation is another ingredient of our study, used to further constrain the parameters of our CIB model, as well as the cosmological ones. The CMB photons propagating freely from the last scattering surface towards us are gravitationally deflected by the large scale distribution of the matter in the Universe. The gravitational lensing leaves imprints on the CMB temperature and polarisation anisotropies. A map of the lensing potential along the line of sight can be reconstructed using these imprints \citep{Okamoto_2003}. Primary sources for this CMB lensing potential are the dark matter halos located between us and the last scattering surface \citep{Lewis_2006} and a strong correlation between the CIB anisotropies and lensing derived projected mass map has been shown to be expected  \citep[e.g.][]{Song_2003} and indeed measured \citep{Planck_ciblensing_2014} with a high signal-to-noise ratio. \\
We calculated the cross-correlation between the CIB and the CMB lensing potential which is given by
\begin{equation} \label{eq:lens}
C_l^{\nu\phi} = \int b_{eff} \bar{j}(\nu,z) \dfrac{3}{l^2}\Omega_m H_0^2 \left(\dfrac{\chi_* - \chi}{\chi_*\chi}\right) P_{lin}(k = l/\chi,z) d\chi \,,
\end{equation}
where $\chi_*$ is the comoving distance to the CMB last scattering surface, $\Omega_m$ is the matter density parameters and $H_0$ is the value of the Hubble parameter today. From this equation, it is seen that $C_l^{\nu,\phi}$ is proportional to $b_{eff}$ whereas $C_l^{CIB}$ is proportional to $b_{eff}^2$. Therefore, using also the \mbox{CIBxCMB} lensing potential measurement in the CIB model likelihood helps us resolve the degeneracy between the evolution of $b_{eff}$ and $\rho_{\textrm{SFR}}$ to some extent.


\subsection{CIBxISW correlation}\label{ssec:2.3} 
The contribution of the ISW effect to the CMB temperature anisotropies is given by:
\begin{equation}\label{eq:tempisw}
	\delta T_{\textrm{ISW}}(\mathbf{\hat{n}}) = \int_{\eta_r}^{\eta_0} d\eta\: e^{-\tau(\eta)} (\mathrm{\dot{\Phi}} - \mathrm{\dot{\Psi}}) [(\eta_0 - \eta)\mathbf{\hat{n}}, \eta].
\end{equation} 
It represents the integration of the time derivative of the $\mathrm{\Phi}$ and $\mathrm{\Psi}$ Newtonian gauge gravitational potentials (with \citealt{Kodama_1984} conventions) over the conformal time $\eta$, with $\eta_r$ being the initial time deep in the radiation era, $\tau(\eta)$ is the optical depth, and dots denote differentiation with respect to the conformal time $\eta$. 

The cross-correlation between the CIB and ISW effect for a given frequency is then:
\begin{equation}
	C^{cr}(\theta_{\mathbf{\hat{n}_1},\mathbf{\hat{n}_2},\nu}) \equiv \langle \delta T_{\textrm{CIB}}(\mathbf{\hat{n}_1},\nu) \delta T_{\textrm{ISW}}(\mathbf{\hat{n}_2}) \rangle \,,
\end{equation}
which becomes, after decomposing in the Legendre series:
\begin{equation}
	C^{cr}(\theta, \nu) = \sum_{l = 2}^{\infty} \dfrac{2l + 1}{4\pi} C_l^{cr}(\nu) P_l(cos(\theta))\,,
\end{equation}
where monopole and dipole are not included. The ISW effect is expected to be dominant on large angular scales ($\ell$<100). At these scales, the Limber approximation is not accurate enough. Following the calculations of \cite{Garriga_2004}, we get the \mbox{CIBxISW} cross power spectrum without Limber approximation as:
\begin{equation}\label{eq:clcr}
	C_l^{cr}(\nu) = 4\pi \dfrac{9}{25} \int \dfrac{dk}{k} \Delta_\mathcal{R}^2 T_l^{\textrm{ISW}}(k) M_l(k,\nu)
\end{equation}
where $\Delta_\mathcal{R}^2$ is the dimensionless primordial power spectrum related to the primordial power spectrum $\mathcal{P_R} \equiv 2\pi^2 \Delta_\mathcal{R}^2/k^3$. The $M_l$ and $T_l^{\textrm{ISW}}$ terms are:
\begin{align}
  M_l(k, \nu) &= c_\mathrm{{\delta \Psi}} \int_{\eta_0}^{\eta_r} d\eta \: j_l(k[\eta - \eta_0]) a(\eta) b_{eff}(\nu) \bar{j}(\nu, \eta) \bar{\delta} (k, \eta) \\
  T_l^{ISW} (k) &= \int_{\eta_0}^{\eta_r} d\eta \: e^{-\tau(\eta)} j_l(k[\eta - \eta_0]) (c_\mathrm{{\Phi \Psi}}\, \dot{\phi} - \mathrm{\dot{\psi}})
\end{align} 
where $j_l$ are spherical Bessel functions, and $\bar{\delta}$, $\mathrm{\phi}$ and $\mathrm{\psi}$ are the time-dependent part of the dark matter density contrast $\delta$, and the Newtonian gauge gravitational potentials $\mathrm{\Phi}$ and $\mathrm{\Psi}$ respectively. The relation between the two coefficients $c_\mathrm{{\Psi \Phi}}$ and $c_\mathrm{{\delta \Psi}}$ for the adiabatic initial conditions is given as \citep{Ma_1995}:
\begin{equation}
	c_\mathrm{\delta \Psi} \equiv \dfrac{\mathrm{\delta}}{\mathrm{\Psi}} = - \dfrac{3}{2}, \: \: c_\mathrm{\Phi \Psi} \equiv \dfrac{\mathrm{\Phi}}{\mathrm{\Psi}} = - \left(1 + \dfrac{2}{5}R_{\nu} \right)
\end{equation}
where $R_\nu \equiv \rho_\nu / (\rho_\nu + \rho_\gamma)$, with $\rho_\nu$ and $\rho_\gamma$ being the energy densities in relativistic neutrinos and photons, respectively. Later in this work, we will need to compute the CIB spectrum at those low multipoles in order to derive the expected covariance of the \mbox{CIBxISW} power spectrum (see details in Sect.~\ref{sec:3}). In order to be as consistent and rigorous as possible, we will use in that context a different expression for the CIB power spectrum compared to Eq.~(\ref{eq:clcib}) -- one that doesn't rely on the Limber approximation:
\begin{equation}
C_l^\mathrm{CIB}(\nu) = 4\pi \dfrac{9}{25} \int \dfrac{dk}{k} \Delta_\mathcal{R}^2\, M_l^2(k,\nu),
\end{equation}

We modified the publicly available Boltzman code {\tt CAMB} to compute our observables of interest: the CIB, \mbox{CIBxISW} and \mbox{CIBxCMB} lensing power spectra, on top of the standard {\tt CAMB} outputs i.e. the CMB temperature and polarisation power spectra.


\begin{table*}[ht]
	\centering
	\begin{tabular}{cccccccc}
		\hline
CIB model & $\alpha$ & $\beta$ & $\gamma$ & $\delta$ & b$_0$ & b$_1$ & b$_2$ \\  
& 0.007 & 3.590 & 2.453 & 6.578 & 0.830 & 0.742 & 0.318 \\ \hline  \hline
		
Cosmology & H$_0$ & $w$ & $\Omega_bh^2$ & $\Omega_ch^2$ & $\tau$ & $n_s$ & $10^9A_s$ \\
		& 67.47 & -1.019 & 0.022 & 0.119 & 0.062 & 0.965 & 2.128 \\ \hline
		
	\end{tabular}
	\caption{Parameter values of the CIB and cosmological models. } \label{tab:pars}
\end{table*}

\section{Expected CIBxISW cross-correlation and significance}\label{sec:3}

Using the formalism described in the previous section, we focus first on revisiting the predictions of \citet{Ilic_2011} for the expected power spectrum and the signal-to-noise ratio of the \mbox{CIBxISW} cross-correlation, using the improved and more tightly constrained CIB model of \citet{Maniyar_2018} mentioned earlier. Before proceeding further, a fiducial set of parameter needs to be agreed upon in order to carry out our computations. For the sake of coherence, we follow \citet{Maniyar_2018}: as the CIB alone cannot constrain cosmological parameters, the authors first assumed a Planck fiducial background cosmology \citep{Planck_cosmo_2016} which we will choose as our fiducial cosmological parameters throughout this work. The authors then derived the best-fit parameters of their CIB model using a combination of multi-frequency measurements of the CIB and \mbox{CIBxCMB} lensing power spectra from Planck and IRAS, as well as star formation rate density measurements and mean CIB intensity levels from external surveys. We use these best-fit CIB parameters as our fiducial values throughout this work. They are given in Table\,\ref{tab:pars}.

\subsection{CIBxISW cross-correlation spectrum}\label{ssec:3.1}

\begin{figure}
\centering
\includegraphics[width=9cm]{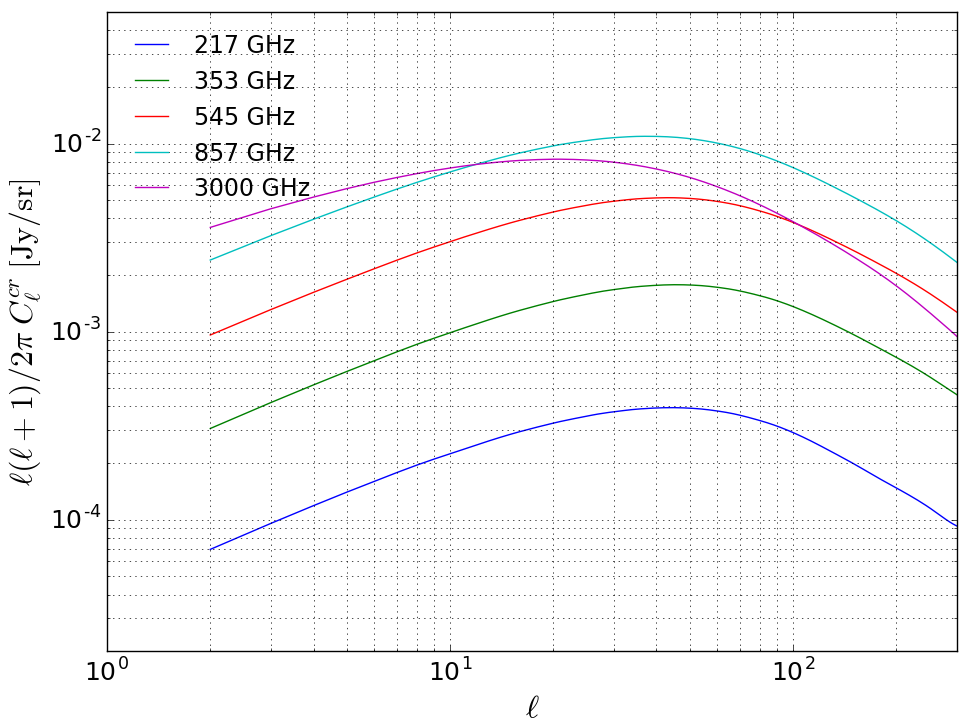}
\centering \caption{CIBxISW cross-correlation power spectra (in Jy\,sr$^{-1}$) for the 5 frequency channels considered in our work.}
\label{fig:cross}
\end{figure}

We compute the \mbox{CIBxISW} cross power spectra for Planck 217, 353, 545, 857\,GHz and IRAS 3000\,GHz frequency channels -- the same frequencies as the CIB datasets used by \citet{Maniyar_2018}. These power spectra are shown in Fig.~\ref{fig:cross}. The amplitude of the cross-correlation is maximal around 20 < $\ell$ < 50 and falls off quickly at higher multipoles as expected from the ISW effect. The amplitude of the power spectra increases with the frequency. This is expected as the CIB probes galaxies at higher redshifts on low frequency channels (and vice versa) and the main contribution to the ISW effect comes from low redshifts. This can be appreciated from the CIB anisotropy redshift distributions provided in Fig.~4 of \cite{Maniyar_2018}, where we see that the distribution peaks towards lower redshifts as we go towards higher frequencies. The amplitude of the signal is the highest for 857\,GHz which is slightly higher than the amplitude at 3000\,GHz. The reason behind this is that the CIB has slightly more power at lower multipoles at 857\,GHz than at 3000\,GHz. We do not consider the signal at higher multipoles ($\ell$ > 100) as the nonlinear counterpart to the ISW effect -- the Rees-Sciama effect -- starts contributing significantly to the CMB anisotropies at those scales. The linear ISW effect however still dominates near the peak of the signal \citep{Seljak_1996}. 

\subsection{Signal-to-noise ratio}\label{ssec:3.2}

Once we calculated the expected \mbox{CIBxISW} cross-correlation signal, we determine the significance (detection level) of the signal. To do so, we perform a signal-to-noise ratio (SNR) analysis. The cumulative signal-to-noise ratio for the ISW effect summed over all the multipoles between 2 $\leq \ell \leq 100$ (where most of the signal resides as shown in Fig.~\ref{fig:cross}) for a given frequency $\nu$ can be written as:
\begin{equation}
	\left[\dfrac{S}{N}\right]^2(\nu) = \sum_{\ell = 2}^{\ell_{max}} \left(\frac{S_\ell(\nu)}{N_\ell(\nu)}\right)^2
\end{equation}
where the signal term $S_\ell$ in our case is the cross-correlation power spectrum:
\begin{equation}\label{eq:signal}
S_\ell(\nu) \equiv C_\ell^{cr}(\nu)\, , 
\end{equation} 
and the noise term $N_\ell$ corresponds to the square root of its variance, the usual cosmic variance:
\begin{equation}\label{eq:noise}
N_\ell(\nu) \equiv \sqrt{\dfrac{\left[C_\ell^{cr}(\nu)\right]^2 + C_\ell^{\mathrm{CIB}}(\nu) C_\ell^{\mathrm{CMB}}}{2\ell + 1}}\, .
\end{equation}
For this analysis, we consider two cases:
\begin{enumerate} 
\item
\textbf{Ideal case}:
we assume that the CIB and the CMB maps used to calculate the cross-correlation are extracted over the full sky and are completely free of residual contaminations (note that instrument noise is not relevant at the angular scales of the ISW effect). The only limiting factor is the cosmic variance and we get the highest SNR in this scenario. With those ideal assumptions, the \mbox{CIBxISW} correlation reaches a significance level as high as $\approx 6.7\sigma$. Detailed results are provided in Table\,\ref{tab:1}. Cumulative SNR for different Planck and IRAS frequencies are shown in Fig.~\ref{fig:cumsnr}. The largest contribution to the SNR comes from $l \leq 50$ after which the significance reaches a plateau. The SNR is the highest for 857 GHz and reaches a value of 6.73. Although the \mbox{CIBxISW} cross-correlation signal is larger for 3000\,GHz frequency than at 217, 353 and 545 GHz at $\ell$ < 100, the SNR is lower than that at these frequencies. This can be explained by looking at the value of the noise given by Eq.~(\ref{eq:noise}). The noise term for 3000\,GHz frequency is much higher at larger scales than it is for other frequencies and hence brings the SNR down. Therefore, in this ideal scenario, the optimal frequency for the ISW detection is 857\,GHz. However, in reality, we have to account for sources of noise and contamination in our analysis as well as reduced sky fractions.

\begin{figure}
	\centering
	\includegraphics[width=8.7cm]{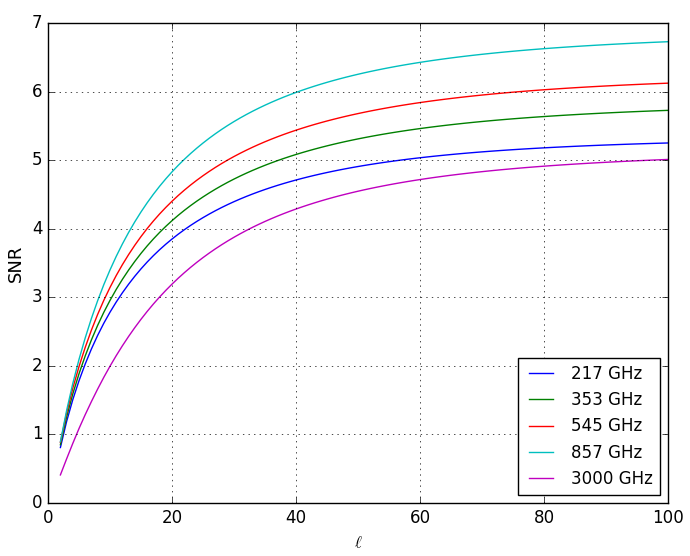}
	\centering \caption{Cumulative signal-to-noise ratio for the \mbox{CIBxISW} cross-correlation signal for the 5 frequency channels considered in our work. The SNR is predicted to be the highest for 857\,GHz. }
	\label{fig:cumsnr}
\end{figure}

\begin{table}[ht]
	\centering
	\begin{tabular}{cccccc}
		\hline
		\hline
		Freq. [GHz]            & 217  & 353  & 545  & 857  & 3000 \\ \hline
		100\% $f_\mathrm{sky}$ & 5.25 & 5.73 & 6.12 & 6.73 & 5.01 \\
		40\% $f_\mathrm{sky}$  & 3.32 & 3.62 & 3.87 & 4.26 & 3.17 \\  \hline
	\end{tabular}
	\newline
	\centering \caption{Predicted SNR for the ISW signal in the ideal scenario (CIB maps free of galactic dust residuals), for 100\% and 40\% sky coverage. The best frequency for the ISW detection in this case is 857\,GHz with SNR reaching 6.73 for full sky and 4.26 for 40\% of the sky. As can be seen from Eq.\,(\ref{eq:snr_real}), the SNR is directly proportional to ${(f_\mathrm{sky})}^{0.5}$ and thus can be easily calculated for other sky fractions in this ideal case.}\label{tab:1}
\end{table}

\item
\textbf{Realistic case:} 
as mentioned earlier, in reality we have to compromise on certain assumptions made in the previous case. In the ideal case, we assumed that the \mbox{CIBxISW} cross-correlation signal can be extracted over the whole sky. On a large part of the sky however, we have strong emissions from our own galaxy. This emission due to the galactic dust is much higher than the CIB signal. This prevents us to extract any CIB signal in regions of the sky too close to the galactic plane. This directly reduces the available sky fraction for measurement by $\sim30$\%. Even then, in the remaining part of the sky the power spectra measurements are contaminated by dust that needs to be removed from the maps. This removal will leave some residual noise that has to be accounted for in the analysis. In addition, as we go to lower frequencies, the CMB starts dominating over the CIB. Therefore, the CMB contribution has also to be removed from the maps. At the scales we are interested in, the CMB is cosmic variance limited and the instrument noise in the CMB measurements can be ignored. When accounting for all those effects, the signal-to-noise ratio then becomes:
\begin{equation}\label{eq:snr_real}
\left[\dfrac{S}{N}\right]^2(\nu) = \sum_{\ell = 2}^{\ell_{max}} (2\ell + 1) \dfrac{f_\mathrm{sky}\,\left[C_\ell^{cr}(\nu)\right]^2 }{\left[C_\ell^{cr}(\nu)\right]^2 + \left[C_\ell^{\mathrm{CIB}}(\nu) + N_\ell^{\mathrm{CIB}}(\nu)\right]\, C_\ell^{\mathrm{CMB}}}
\end{equation}
where the $N_\ell^\mathrm{CIB}(\nu)$ term contains the dust residuals left in the CIB maps and $f_\mathrm{sky}$ represents the fraction of the sky where CMB and CIB are both available. The presence of dust residuals does not affect the measurement of the \mbox{CIBxCMB} cross-correlation as they are not correlated with the CMB. \\
To quantify the dust residuals in the CIB i.e. $N_\ell^\mathrm{CIB}(\nu)$, we used the following procedure:
\begin{itemize}
  \item We computed the dust power spectra using the dust map provided by the {\tt PLA}\footnote{\url{https://pla.esac.esa.int/pla/}} (Planck Legacy Archive) generated using the COMMANDER component separation code \citep{Planck_2016}. This map is provided at the reference frequency of 545\,GHz. In order to account for the available sky fractions, we applied different masks to this map, also provided by {\tt PLA}. We performed our analysis on 20\%, 40\%, 60\%, 70\% and 80\% of the sky. We also considered the 10\% mask provided in Table 2 of \cite {Planck_cib_2014}. After applying these masks, we computed the dust power spectrum from the dust map using the {\tt Xpol} software which is an extension to the polarisation of the {\tt Xspect} method \citep{Tristram_2005}. The analysis on the maps is done using the {\tt HEALPix} package \citep{Healpix_2005}.  
  \item The power spectrum of the dust $C_\ell^{\mathrm{dust}}$ is frequency dependent. Once we obtained the power spectrum of the dust at 545\,GHz, we converted it to other frequencies assuming a modified black body shape for dust emission with a given average dust temperature <T$_d$> and spectral index <$\beta_d$>. The values for <T$_d$> and <$\beta_d$> are taken from Table~3 of \cite{Planck_dust_2014} for $|b|>15^o$ and are <T$_d$> = 20.3\,K and <$\beta_d$> = 1.59. 
  \item We briefly explain in Sect.~\ref{ssec:5.1} that the HI gas column density is observed to be proportional to the galactic dust emission. This relation is used to clean the CIB maps from dust contamination. Dust residuals left in the CIB maps using the HI gas as a dust tracer has been quantified by \cite{Planck_cib_2014}, as detailed in their Sect.\,3.2.2 and 3.2.3. They simulate multiple maps of the dust emissivity and obtain the corresponding dust map for a given frequency by multiplying the dust emissivity maps by the GASS map of HI column densities. These dust maps are then used to create simulated maps of total sky emission at all frequencies (e.g., including CIB). Using these sky maps and performing the cleaning of the dust through the above mentioned galactic dust-HI gas column density relation, \cite{Planck_cib_2014} find the dust residuals on the power spectrum to be of the order of 5-10\% for all the frequencies (5\% for 217\,GHz and 10\% at 857\,GHz). Thus, and to be conservative, we considered here a residual dust power spectrum of 10\% of $C_\ell^{\mathrm{dust}}(\nu)$ at all the frequencies. Therefore, $N_\ell^\mathrm{CIB}(\nu) = 0.1 \times C_\ell^{\mathrm{dust}}(\nu)$.
\end{itemize}
After calculating the dust power spectra and eventually $N_\ell^\mathrm{CIB}(\nu)$, we can calculate the new SNR using 
Eq.~(\ref{eq:snr_real}). The results are presented in Table~\ref{tab:2}. We can see that the SNR decreases quite dramatically once the dust residuals are included in the analysis. Also, the SNR decreases when increasing the available sky fraction for the analysis, which is the exactly opposite to the ideal dust free scenario where higher sky fraction resulted in higher SNR. The impact of the dust on the CIB power spectra can be seen from Fig.~\ref{fig:cibanddust} where we plot the CIB and the 10\% residual dust power spectra at 353 and 3000\,GHz. The residual dust power spectra are calculated over 20, 40, 60, 70 and 80\% of the sky. It can be seen that at the lower multipoles i.e. $\ell$ < 100 where we expect the ISW signal to be detected, the CIB spectra are completely dominated by the dust residuals. This significantly impacts the SNR of the ISW signal and brings it down. The highest SNR is obtained for 353\,GHz with 20\% of the sky fraction available for the analysis. We can see from Eq.~(\ref{eq:snr_real}) that a smaller sky fraction results in decreasing the SNR for the ISW detection in the case of dust-free spectra. At the same time, it's clear from Fig.~\ref{fig:cibanddust} that the residual dust power comes down with smaller sky fractions and hence increases the SNR. Therefore, there is a trade-off between the available sky fraction $f_\mathrm{sky}$ and the SNR for all the frequencies and in this case, a 20\% sky fraction provides the best SNR for 353\,GHz. 

\begin{figure*}[ht!]
\centering
\includegraphics[width=18cm,height = 10cm]{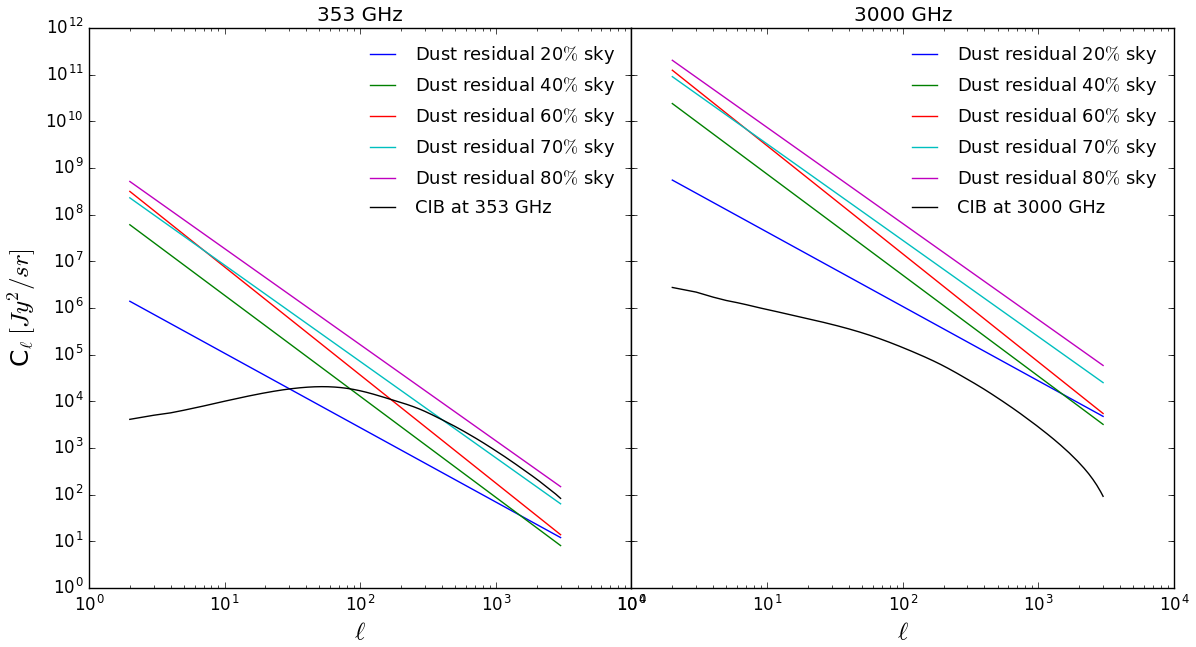}
\centering \caption{Power spectra of the CIB and 10\% of the dust residual at 353\,GHz (left panel) and 3000\,GHz (right panel). The dust power spectra
are computed over 20, 40, 60, 70 and 80\% of the sky. We see that the residual dust power spectra dominate over the CIB at the low multipoles of interest for the ISW signal. As a consequence, the SNR decreases dramatically compared to the ideal dust free case.}
\label{fig:cibanddust}
\end{figure*}

\begin{table}[ht]
 \centering
\begin{tabular}{cccccc}
\hline
\hline 
Freq. [GHz] &  217 & 353 & 545 & 857 & 3000\\ \hline
10\% $f_\mathrm{sky}$  & 0.88 & 0.91 & 0.83 & 0.65 & 0.28 \\
20\% $f_\mathrm{sky}$  & 1.46 & 1.51 & 1.41 & 1.14 & 0.52 \\
40\% $f_\mathrm{sky}$   & 1.16 & 1.16 & 0.99 & 0.69 & 0.27 \\
60\% $f_\mathrm{sky}$   & 0.92 & 0.90 & 0.74 & 0.49 & 0.18 \\
70\% $f_\mathrm{sky}$ & 0.80 & 0.78 & 0.63 & 0.42 & 0.16 \\
80\% $f_\mathrm{sky}$ & 0.59 & 0.57 & 0.45 & 0.30 & 0.11 \\ \hline
\end{tabular}
\newline
\centering \caption{Predicted SNR for the ISW signal in the current realistic scenario with the inclusion of the dust residuals in the CIB maps. They are given for 10, 20, 40, 60, 70 and 80\% of the sky. These SNR have been calculated assuming that there is a 10\% residual dust in the CIB maps. It is seen that we basically detect no signal with a good SNR for the ISW effect in these cases. The best case scenario seems to be for the 353\,GHz frequency over 20\% sky where SNR reaches 1.51.}\label{tab:2}
\end{table}
\end{enumerate}


\section{Forecasting constraints on cosmological parameters}\label{sec:4}

We showed in Sect.~\ref{ssec:3.2} that in the ideal scenario with the dust free CIB and CMB maps, the \mbox{CIBxISW} cross-correlation can provide us with the detection of the ISW effect with the highest SNR for a single tracer. Unfortunately, we are currently limited by the dust residuals in the CIB which reduces our ability to measure the ISW signal with a good SNR. However in the near future, improved cleaning techniques, additional datasets and a better understanding of contaminants could allow us to improve the CIB maps, with much lower levels of dust residuals. In such conditions, the \mbox{CIBxISW} cross-correlation would again become a very competitive probe of cosmology and dark energy in particular. Exploiting the \mbox{CIBxCMB} lensing cross-correlation as well could push the constraints on cosmological parameters even further. With the prospect of those future improved datasets in mind, we perform in this section a joint Fisher matrix analysis including the CMB, the CIB, the \mbox{CIBxCMB} lensing cross-correlation and the ISW effect through the \mbox{CIBxCMB} cross-correlation and ignoring dust residuals in the CIB.


\subsection{Fisher matrix formalism}\label{ssec:4.1}

We recall here briefly the principles underlying the Fisher matrix formalism. Consider a set of hypothetical $N$ Gaussian data $\mathbf{x}$ with mean $\boldsymbol{\mu}$ and covariance matrix $\boldsymbol{\Sigma}$ distributed according to the likelihood:
\begin{equation}
	L = \frac{1}{(2\pi)^{n/2}\sqrt{\det \boldsymbol{\Sigma}}} \exp\left[-\frac{1}{2}(\mathbf{x}-\boldsymbol{\mu})^T \boldsymbol{\Sigma}^{-1} (\mathbf{x}-\boldsymbol{\mu}) \right] \,.
\end{equation}
Here both $\boldsymbol{\mu}$ and $\boldsymbol{\Sigma}$ depend on $\boldsymbol{\theta} = \left[ \theta_1, \theta_2,  \dots \right ]$, the vector of parameters of the model we try to constrain. We are able to forecast the constraints on those parameters (that the hypothetical dataset would provide) in the form of a covariance matrix $\mathbf{C}$, obtained by inverting the so-called Fisher matrix $\mathbf{F}$. The elements of this matrix are defined as the following expectation value:
\begin{equation}
	F_{m, n} = \left \langle  \frac{\partial^2 (-\ln L)}{\partial \theta_m \partial \theta_n} \right \rangle
\end{equation}
which after some work can be rewritten as:
\begin{equation} \label{eq:Fisher}
	F_{m, n} = \dfrac{\partial \boldsymbol{\mu}}{\partial \theta_m} \boldsymbol{\Sigma}^{-1} \dfrac{\partial \boldsymbol{\mu}^T}{\partial \theta_n} + \dfrac{1}{2} \textrm{tr} \left(\boldsymbol{\Sigma}^{-1} \dfrac{\partial \boldsymbol{\Sigma}}{\partial \theta_m} \boldsymbol{\Sigma}^{-1} \dfrac{\partial \boldsymbol{\Sigma}}{\partial \theta_m} \right),
\end{equation}
where $\textrm{tr}()$ denotes the trace of a square matrix.

In our work, we will fix the covariance matrix to its fiducial value (a common assumption in Fisher analyses) and hence the second term in the Eq.~(\ref{eq:Fisher}) vanishes as $\dfrac{\partial \boldsymbol{\Sigma}}{\partial \theta_m} = 0$ and therefore:
\begin{equation}\label{eq:fish}
	F_{m, n} = \dfrac{\partial \boldsymbol{\mu}}{\partial \theta_m} \boldsymbol{\Sigma}^{-1} \dfrac{\partial \boldsymbol{\mu}^T}{\partial \theta_n}\,.
\end{equation}
When needed, we compute the partial derivatives as
\begin{equation}\label{eq:dmudtheta}
	\dfrac{\partial\boldsymbol{\mu}}{\partial\theta_m} = \dfrac{\boldsymbol{\mu}(\boldsymbol{\theta'}, \theta_m + \delta\theta_m) - \boldsymbol{\mu}(\boldsymbol{\theta'}, \theta_m - \delta\theta_m)}{2\,\delta\theta_m} \,,
\end{equation}
where for a given parameter $\theta_m$, the vector $\boldsymbol{\theta'}$ represent all the other parameters kept constant at their fiducial value. The value of $\delta\theta$ is chosen small enough to stay in the Gaussian regime of the likelihood, but also large enough not to be affected by the numerical errors.

\medskip

More practically, in our analysis, the parameters of the model are:
\begin{equation}
	\boldsymbol{\theta} = \left[ \boldsymbol{\theta}_{\rm CIB}, \boldsymbol{\theta}_{\rm cosmo.} \right]
\end{equation}
where: 
\begin{equation} 
	\boldsymbol{\theta}_{\rm CIB} = \left[ \alpha, \beta, \gamma, \delta, b_0, b_1, b_2, f_{\nu_i}^\textrm{cal} \right]
\end{equation}
are the 7 CIB model parameters, plus 5 calibration factors $f_{\nu_i}^\textrm{cal}$ (one for each frequency) which account for relative absolute calibration uncertainties between the different frequency channels. On the other hand:
\begin{equation}
	\boldsymbol{\theta}_{\rm cosmo.} = \left[ H_0,  \Omega_bh^2, \Omega_ch^2, \tau, n_s, A_s \right]
\end{equation}
are the $\Lambda$CDM model parameters, to which we add $w$ when we consider the $w$CDM model. Then, in our case the vector $\boldsymbol{\mu}(\boldsymbol{\theta})$ contains the predicted values of our $N$ observables of interest as a function of our model parameters, and $\boldsymbol{\Sigma}(\boldsymbol{\theta})$ is their theoretical covariance matrix. 

As mentioned before, we consider the four following probes as observables: the CMB power spectra, the CIB power spectra ($+$ external constraints), the \mbox{CIBxCMB} lensing cross-spectra and the \mbox{CIBxISW} cross-spectra. Our choices for the properties of those observables allow us to simplify the Fisher analysis. Indeed, we use the CMB as a prior in our work, and thus neglect its correlation with the other three probes. Also, we define our CIB and \mbox{CIBxCMB} lensing power spectra similarly to the actual data used in \citet{Maniyar_2018}, meaning that the spectra are ``binned'' in multipoles\footnote{More precisely: we consider 3 multipole bins for the CIB power spectra between $\ell \in[145, 592]$ and 15 bins for the \mbox{CIBxCMB} lensing spectra between $\ell \in [100, 2000].$}. This allows us firstly to ignore the correlations between bins for a given probe, and secondly to neglect correlations between the probes as their respective bins barely overlap. Finally, we can also safely ignore the correlation between the \mbox{CIBxISW} spectra and the previous two, as their multipole range do not overlap ($\ell \in [2, 100]$). Since we do not need to consider the correlations between our different probes, we can write our total Fisher matrix as the sum of the individual Fisher matrices of each probe:
\begin{equation}
	\mathbf{F} = \mathbf{F}_{\rm CMB} + \mathbf{F}_{\rm CIB\,+\,ext.} + \mathbf{F}_{\rm CIB\,x\,CMB\,lensing} + \mathbf{F}_{\rm CIB\,x\,ISW}
\end{equation}
In the following sections, we explain the construction of the Fisher matrices of the individual probes.


\subsubsection{CMB Fisher matrix}\label{ssec:4.1.1}

Our aim is to explore the relative improvement on the constraints on the cosmological parameters set by the CMB using the ISW signal detected through the \mbox{CIBxCMB} cross-correlation. For this reason, we treat the CMB as prior information on all cosmological parameters and use the latest Planck data (plus some external data in the $w$CDM case, see Sect.~\ref{ssec:4.3}) as constraints.
Instead of computing the theoretical $\boldsymbol{\mu}(\boldsymbol{\theta})$ and $\boldsymbol{\Sigma}(\boldsymbol{\theta})$ for the CMB power spectra, we directly derive an ``equivalent'' CMB Fisher matrix from Planck data using the official MCMC chains publicly available on the {\tt PLA}\footnote{We use the {\tt 'plikHM\_TTTEEE\_lowTEB'} chains for the $\Lambda$CDM model and the {\tt 'plikHM\_TTTEEE\_lowTEB\_BAO\_H070p6\_JLA'} ones for the $w$CDM model.} by computing their covariance matrix (using the GetDist package provided by Antony Lewis) and inverting it.
Note that since the CMB does not provide any information about the CIB model parameters, the corresponding rows and columns of the CMB Fisher matrix are set to 0.


\subsubsection{CIB-related Fisher matrix}\label{ssec:4.1.2}

On top of the CMB prior, we then add as much information as possible to constrain our CIB model parameters. To do so, we consider for our observable vector $\boldsymbol{\mu}$ the same CIB probes as those used as data in \citet{Maniyar_2018}. These include the CIB auto and cross power spectra, mean level of the CIB at different frequencies, star formation rate measurements, and the \mbox{CIBxCMB} lensing cross-correlation. The theoretical value of all those observables for any set of parameters can be computed using the models described earlier in Sect.~\ref{ssec:2.1} and \ref{ssec:2.2}. Their derivatives with respect to our model parameters are computed using Eq.~(\ref{eq:dmudtheta}).

For those observables, instead of computing an analytical covariance matrix $\boldsymbol{\Sigma}$ -- that may lack subtle features present in the real signal (effects of the foreground removals, partial sky coverage, etc.) -- we make use of the available CIB data and use directly the estimated covariance matrix of the data from \citet{Planck_cib_2014} as our fiducial $\boldsymbol{\Sigma}$. As mentioned at the beginning of Sect.~\ref{ssec:4.1}, all CIB observables are considered independent therefore only the diagonal blocks of the covariance matrix $\boldsymbol{\Sigma}$ will be non-zero. This allows us to compute their respective Fisher matrices independently -- using Eq.~(\ref{eq:Fisher}) -- and simply add them together afterwards. We consider two hypothetical scenarios: one optimistic scenario where we obtained a CIB map over $f_{\rm sky}$~=~$70\%$ of the sky, and a more pessimistic one with $f_{\rm sky}$~=~$40\%$. We multiply the $\boldsymbol{\Sigma}$ data covariance from \citet{Planck_cib_2014} -- which was obtained over 10\% of the sky -- by a factor of $0.1/f_{\rm sky}$ to account for this hypothetical improvement of the CIB recovery.


\subsubsection{CIBxISW Fisher matrix}\label{ssec:4.1.3}

We consider here as observables the \mbox{CIBxISW} cross-correlation power spectra, for all 5 observed CIB frequencies of \citet{Maniyar_2018} and multipoles between $\ell \in [2, 100]$. To emulate a future analysis of actual data, and similarly to the CIB power spectra, we choose here to work with binned \mbox{CIBxISW} cross-correlation power spectra, splitting the multipole range into 11 roughly equal bins. For a given $[\ell_1, \ell_2]$ bin, we define the binned spectrum $B_{\ell_1,\ell_2}$ as a simple arithmetic mean:
\begin{equation}\label{eq:binnedcl}
	B_{\ell_1, \ell_2} = \frac{1}{n_\ell} \sum_{\ell \in [\ell_1, \ell_2]} C_\ell
\end{equation}
where $n_\ell$ is the number of multipoles in the bin, and $C_\ell$ in this case is the \mbox{CIBxISW} power spectrum as computed using the model described in Sect.~\ref{ssec:2.3} and their derivative using Eq.~(\ref{eq:dmudtheta}).

We compute the covariance of the \mbox{CIBxISW} power spectrum using a generalization of Eq.~(\ref{eq:noise}); the covariance between the cross-spectrum $C^{cr}_\ell(\nu_1)$ at frequency $\nu_1$ and $C^{cr}_\ell(\nu_2)$ at $\nu_2$ is written as:
\begin{equation}\label{eq:binnedcov}
	{\rm Covar}(C^{cr}_\ell(\nu_1), C^{cr}_\ell(\nu_2)) = \frac{C_\ell^{cr}(\nu_1)C_\ell^{cr}(\nu_2) + C_\ell^{\mathrm{CIB}}(\nu_1,\nu_2) C_\ell^{\mathrm{CMB}}}{f_{\rm sky} (2\ell+1)}
\end{equation}
Similarly to the CIB observables, we consider an optimistic scenario with a dust-free CIB map available over $f_{\rm sky}$~=~$70\%$ of the sky, and a more pessimistic one with $f_{\rm sky}$~=~$40\%$. Combining Eq.~(\ref{eq:binnedcov}) with the definition of Eq.~(\ref{eq:binnedcl}), we can derive the covariance between the binned cross-correlation $B^{cr}_{\ell_1, \ell_2}$ at two frequencies:
\begin{equation}
	{\rm Covar}(B^{cr}_{\ell_1, \ell_2}(\nu_1), B^{cr}_{\ell_1, \ell_2}(\nu_2)) = \frac{1}{n_\ell^2}\sum_{\ell \in [\ell_1, \ell_2]} {\rm Covar}(C^{cr}_\ell(\nu_1), C^{cr}_\ell(\nu_2))
\end{equation}
which allows us to get the covariance matrix $\boldsymbol{\Sigma}$ required to compute the \mbox{CIBxISW} Fisher matrix, again using Eq.~(\ref{eq:Fisher}).


\subsection{Forecasts for the $\Lambda$CDM model}\label{ssec:4.2}

\begin{table*}[ht]
	\centering
	\begingroup
	\setlength{\tabcolsep}{3.5pt}
	\renewcommand*{\arraystretch}{1.4}
	\begin{tabular}{lllcccccc|c}
		\hline
		\multicolumn{10}{c}{$\Lambda$CDM model, $f_{\rm sky}=70\%$ case} \\
		\hline
		& & & $\sigma(H_0)$ & $\sigma(\Omega_bh^2)$ & $\sigma(\Omega_ch^2)$ & $\sigma(\tau)$ & $\sigma(n_s)$ & $\sigma(10^9A_s)$ & $\sigma(\Omega_\Lambda)$ \\
		CMB + CIB & & & 0.344 & 0.000121 & 0.000812 & 0.0162 & 0.00389 & 0.0710 & 0.00474 \\
		 & + CIBxCMB lensing & & 0.574\% & 0.535\% & 0.496\% & 33.9\% & 1.73\% & 36.4\% & 0.538\% \\
		 & & + CIBxISW & 19.2\% & 39.0\% & 17.0\% & 38.7\% & 8.05\% & 34.5\% & 17.4\%\\
		\multicolumn{10}{c}{\ } \\
		\hline
		\multicolumn{10}{c}{$\Lambda$CDM model, $f_{\rm sky}=40\%$ case}\\
		\hline
		& & & $\sigma(H_0)$ & $\sigma(\Omega_bh^2)$ & $\sigma(\Omega_ch^2)$ & $\sigma(\tau)$ & $\sigma(n_s)$ & $\sigma(10^9A_s)$ & $\sigma(\Omega_\Lambda)$ \\
		CMB + CIB & & & 0.412 & 0.000128 & 0.000950 & 0.0164 & 0.00410 & 0.0715 & 0.00565 \\
		 & + CIBxCMB lensing & & 0.485\% & 0.410\% & 0.458\% & 24.4\% & 1.24\% & 26.3\% &  0.474\% \\
		 & & + CIBxISW & 18.1\% & 31.5\% & 15.3\% & 34.2\% & 8.19\% & 31.3\% & 16.4\%\\
		\multicolumn{10}{c}{\ }
	\end{tabular}
	\endgroup
	\caption{Forecasted constraints on $\Lambda$CDM cosmological parameters. The upper and lower tables show respectively our results in the $f_{\rm sky}=70\%$ and $f_{\rm sky}=40\%$ cases. In each table, the first line shows the predicted 1$\sigma$ errors from CMB\,+\,CIB constraints. The second one shows the improvements on those errors (i.e. relative reduction in \%) after adding the \mbox{CIBxCMB} lensing constraints. The last line shows the further improvement thanks to \mbox{CIBxISW} constraints, using the \{CMB\,+\,CIB\,+\mbox{CIBxCMB} lensing\} case as a reference. Note that the constraints on $\Omega_\Lambda$ are derived from the other parameters (and not originally part of the Fisher analysis). }\label{tab:lcdm}
\end{table*}

\begin{figure*}[ht]
	\centering
	\includegraphics[width=\textwidth]{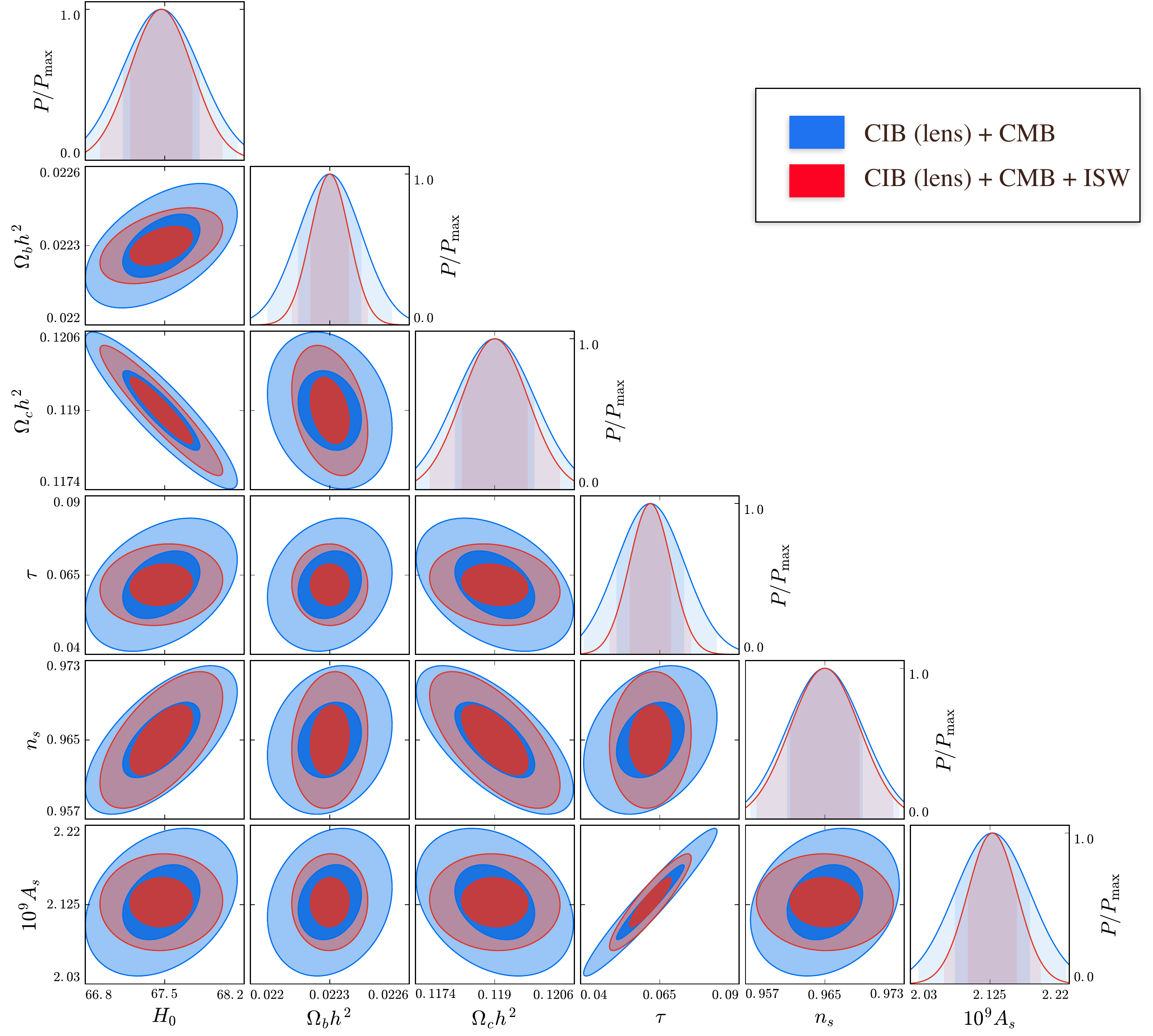}
	\caption{Fisher forecasts for the 1$\sigma$ and 2$\sigma$ confidence contours for the $\Lambda$CDM cosmological parameters in the $f_{\rm sky}=70\%$ case. The blue contours show our results when the CMB, CIB and \mbox{CIBxCMB} lensing constraints are combined. The red coloured show the same results when adding the \mbox{CIBxISW} cross-correlation constraints. There is a $\sim17$\% improvement on the constraints on $\Omega_\Lambda$ and $\sim39/35$\% for $\tau$/$A_s$ which are strongly correlated.}
	\label{fig:lcdm}
\end{figure*}

We first present our Fisher forecasts in the context of the $\Lambda$CDM model. Table~\ref{tab:lcdm} shows the 1$\sigma$ constraints on cosmological parameters in the various scenarios we explored, separating our results for our two choices of $f_{\rm sky}$. We show those constraints in the \{CMB\,+\,CIB\} case, and their relative improvements due to the addition of the \mbox{CIBxCMB} lensing alone, and the further improvements after adding the \mbox{CIBxISW} constraints. Fig.~\ref{fig:lcdm} gives a visual representation of some of those constraints. These contours have been generated using the {\tt CosmicFish} software \citep{Cosmicfish_2016}. The blue coloured ellipses represent the forecasted 1$\sigma$ and 2$\sigma$ confidence contours obtained from the combination of CMB, CIB and \mbox{CIBxCMB} lensing information. The red coloured ellipses are the corresponding confidence regions after adding the constraints from the \mbox{CIBxISW} signal.

As mentioned before, the ISW effect is a probe of the growth of structures through time (see Eq.~(\ref{eq:tempisw})), which itself is strongly dependent on the characteristics of the dark energy model considered. The cosmological constant $\Lambda$ affects cosmology only at fairly late times; coupled with the fact that its energy density has remained low and constant in time, the ISW effect will not show its full potential as a probe in the flat $\Lambda$CDM model.  The two main parameters governing the growth of structures in the flat $\Lambda$CDM model are the dark energy density $\Omega_\Lambda$ (or equivalently the matter density $\Omega_m=1-\Omega_\Lambda$), and the amplitude of matter fluctuations through the power spectrum amplitude $A_s$. We thus expect the constraints on those two parameters to be the most improved after adding the ISW signal in the Fisher analysis \citep[see also][]{Coble_1997}. This expectation is indeed verified: using the \{CMB\,+\,CIB\,+\mbox{CIBxCMB} lensing\} case as a reference, we obtain an improvement of $\sim$17\% on the error bars on $\Omega_\Lambda$ for $f_{\rm sky}=70\%$ ($\sim$16\% for $f_{\rm sky}=40\%$) and of $\sim35$\% ($\sim31$\%) on $A_s$ when the \mbox{CIBxISW} information is added in the Fisher matrix (illustrated in Fig.~\ref{fig:lcdm}). There is a well known degeneracy between $\tau$ and $A_s$ \citep[e.g.][]{Planck_cosmo_2016} which benefits from the tightening of the $A_s$ constraints: we thus obtain a similar improvement of $\sim39$\% ($\sim34$\%) on the $\tau$ parameter after adding the ISW information. 

Interestingly, Table~\ref{tab:lcdm} shows that the relative improvements in all constraints when adding \mbox{CIBxCMB} lensing and ISW information is roughly the same for both $f_{\rm sky}$ (although the absolute constraints are of course better in the 70\% case.) 

We note that the addition of the \mbox{CIBxCMB} lensing cross-correlation alone also improves the constraints on some cosmological parameters. As the CMB lensing signal probes the distribution of matter while being independent of $\tau$ \citep{Planck_overview_2014}, we expect it to help break the degeneracy between $\tau$ and $A_s$. This is indeed the case as we see from Table~\ref{tab:lcdm} that adding the \mbox{CIBxCMB} lensing cross-correlation mainly improves the constraints on those two parameters. 

\subsection{Forecasts for the $w$CDM model}\label{ssec:4.3}

\begin{figure*}[ht]
	\centering
	\includegraphics[width=\textwidth]{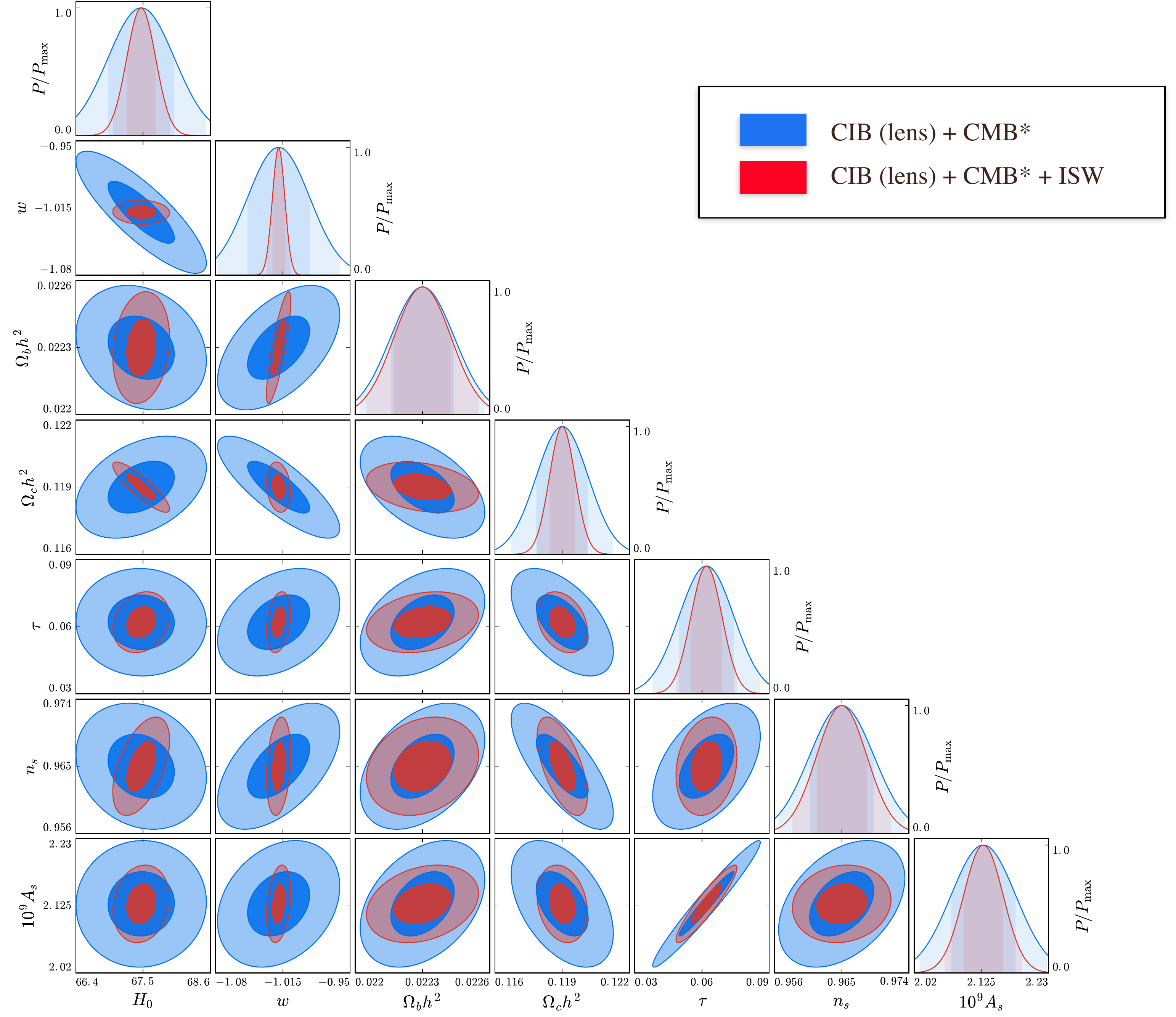}
	\caption{Fisher forecasts for the 1$\sigma$ and 2$\sigma$ confidence contours for the $w$CDM cosmological parameters in the $f_{\rm sky}=70\%$ case. The blue contours show our results when the CIB, \mbox{CIBxCMB} lensing, and the CMB along with the external (JLA + BAO and $H_0$ prior) constraints on the cosmological parameters constraints are combined. The red coloured show the same results when adding the \mbox{CIBxISW} cross-correlation constraints. There is a $\sim80$\% improvement on the constraints on $w$, $\sim29$\% for $\Omega_\Lambda$, and $\sim43/39$\% for $\tau$/$A_s$ which are strongly correlated.}
	\label{fig:wcdm}
\end{figure*}

\begin{table*}
	\centering
	\begingroup
	\setlength{\tabcolsep}{3.5pt}
	\renewcommand*{\arraystretch}{1.4}
	\begin{tabular}{lllccccccc|c}
		\hline
		\multicolumn{11}{c}{$w$CDM model, $f_{\rm sky}=70\%$ case} \\
		\hline
		& & & $\sigma(H_0)$ & $\sigma(w)$ & $\sigma(\Omega_bh^2)$ & $\sigma(\Omega_ch^2)$ & $\sigma(\tau)$ & $\sigma(n_s)$ & $\sigma(10^9A_s)$ & $\sigma(\Omega_\Lambda)$ \\
		CMB* + CIB & & & 0.561 & 0.0308 & 0.000143 & 0.00118 & 0.0170 & 0.00434 & 0.0730 & 0.00480\\
		 & + CIBxCMB lensing & & 2.98\% & 2.20\% & 0.577\% & 1.13\% & 27.7\% & 0.756\% & 30.7\% & 2.93\% \\
		 & & + CIBxISW & 56.4\% & 80.1\% & 10.1\% & 50.8\% & 43.4\% & 22.0\% & 38.7\% & 29.4\%\\
		\multicolumn{11}{c}{\ } \\
		\hline
		\multicolumn{9}{c}{$w$CDM model, $f_{\rm sky}=40\%$ case}\\
		\hline
		& & & $\sigma(H_0)$ & $\sigma(w)$ & $\sigma(\Omega_bh^2)$ & $\sigma(\Omega_ch^2)$ & $\sigma(\tau)$ & $\sigma(n_s)$ & $\sigma(10^9A_s)$ & $\sigma(\Omega_\Lambda)$ \\
		CMB* + CIB & & & 0.656 & 0.0330 & 0.000144 & 0.00121 & 0.0170 & 0.00437 & 0.0730 & 0.00575 \\
		 & + CIBxCMB lensing & & 2.91\% & 1.85\% & 0.322\% & 0.749\% & 20.2\% & 0.553\% & 22.3\% & 3.15\%\\
		 & & + CIBxISW & 55.4\% & 79.5\% & 9.15\% & 43.6\% & 36.5\% & 18.8\% & 33.2\% & 29.8\%\\
		\multicolumn{11}{c}{\ } \\
	\end{tabular}
	\endgroup
	\caption{Forecasted constraints on $w$CDM cosmological parameters. The upper and lower tables show respectively our results in the $f_{\rm sky}=70\%$ and $f_{\rm sky}=40\%$ cases. In each table, the first line shows the predicted 1$\sigma$ errors from CMB*\,+\,CIB constraints. The second one shows the improvements on those errors (i.e. relative reduction in \%) after adding the \mbox{CIBxCMB} lensing constraints. The last line shows the further improvement thanks to \mbox{CIBxISW} constraints, using the \{CMB\,+\,CIB\,+CMBxCIB lensing\} case as a reference. Note that the constraints on $\Omega_\Lambda$ are derived from the other parameters (and not originally part of the Fisher analysis).}\label{tab:wcdm}
\end{table*}

As pointed out by various studies \citep[e.g.][]{Coble_1997,Corasaniti_2003}, the ISW effect has a strong dependence on the equation of state $w\equiv P/\rho$ of the dark energy. This is expected, as any deviation from a pure cosmological constant (i.e. $w\neq-1$) has a strong influence on the growth rate of structures at all times, which in turn affect the time variation of the potential wells that the ISW effect directly probes. Combined with the fact that the CIB signal at different frequencies probes the galaxy distribution (and thus the growth of structure) at different redshifts, we thus expect that the addition of the CIB-ISW signal in our Fisher analysis will improve the constraints on $w$CDM parameters much more than it did for $\Lambda$CDM. We performed a Fisher analysis for the $w$CDM model similar to the case of the $\Lambda$CDM model. Here, along with the 6 cosmological parameters of the $\Lambda$CDM model, we have the additional parameter $w$.

There is a well-known geometrical degeneracy between $w$ and the Hubble constant $H_0$ in the context of CMB studies \citep[e.g.][]{Planck_cosmo_2016} in order to preserve the angular size of the sound horizon at the last scattering surface. Therefore, the CMB data alone cannot constrain both $H_0$ and $w$ together. External cosmological data sets constraining the background evolution of the Universe, such as baryonic acoustic oscillations (BAO), supernovae type Ia (SNe Ia), and local $H_0$ measurements, are required to break this degeneracy. In order to extract a valid covariance matrix (and thus deduce an equivalent Fisher matrix) directly from the Planck MCMC chains, the corresponding posterior constraints on the cosmological parameters need to follow a Gaussian distribution. This is not the case for the CMB-only chains, due to the aforementioned $w$-$H_0$ degeneracy. Therefore, we used the Planck chains provided on the {\tt PLA} where the CMB data is combined with several BAO datasets (6dFGS, \citealt{2011MNRAS.416.3017B}, SDSS MGS, \citealt{2015MNRAS.449..835R}, BOSS LOWZ and CMASS, \citealt{2014MNRAS.441...24A}), the JLA supernovae dataset \citep{2014A&A...568A..22B}, and the local $H_0$ measurement from \citet{2014MNRAS.440.1138E}. We refer to the sum of those constraints as CMB$^*$ in the following.

Results for the analysis of the $w$CDM model are presented in Fig.~\ref{fig:wcdm} and Table~\ref{tab:wcdm}. Similarly to the $\Lambda$CDM analysis, the blue and red contours of Fig.~\ref{fig:wcdm} represent our forecasts on cosmological parameters respectively without and with CIB-ISW constraints in the Fisher matrix. We observe a strong improvement on the equation of state $w$ of about 80\% (irrespective of $f_{\rm sky}$) when the ISW signal information is added in the Fisher matrix on top of the CMB, CIB and \mbox{CIBxCMB} lensing information. As $w$ is correlated to other cosmological parameters, improvements are partially propagated. This is especially noticeable for $H_0$ where an improvement of $\sim$56\% is seen for both $f_{\rm sky}$. Similarly to the $\Lambda$CDM case (cf. Sect.~\ref{ssec:4.2}), the ISW signal also improves the constraints on $\Omega_\Lambda$ ($\sim30$\%), $A_s$ ($\sim$39/33\%) and thus $\tau$ ($\sim$43/37\%).

These results illustrate for a simple dark energy model the great potential of the \mbox{CMBxCIB} correlation as a probe to constrain such a model, thanks to the discriminating power of the ISW effect. We can expect it to be of similar potential when constraining further extensions to the standard $\Lambda$CDM model.


\section{Measurement of the ISW effect using the Planck maps}\label{sec:5}

The results shown in the previous sections were obtained assuming more or less idealized scenarios. As seen in Sect.~\ref{ssec:3.2}, including the effects of the dust residuals can significantly lower the SNR. Indeed, in practice the currently available CIB maps contain significant amount of such residuals and therefore, the error bars on the CIB$\times$CMB signal would be dominated by them, cf. Eq.~(\ref{eq:snr_real}). In this section, we present the constraints we can obtain on the cross-correlation measurement using the current CIB and CMB maps.


\subsection{CIB and CMB Planck maps}\label{ssec:5.1}
\begin{figure}[h]
\centering
\includegraphics[width=9cm]{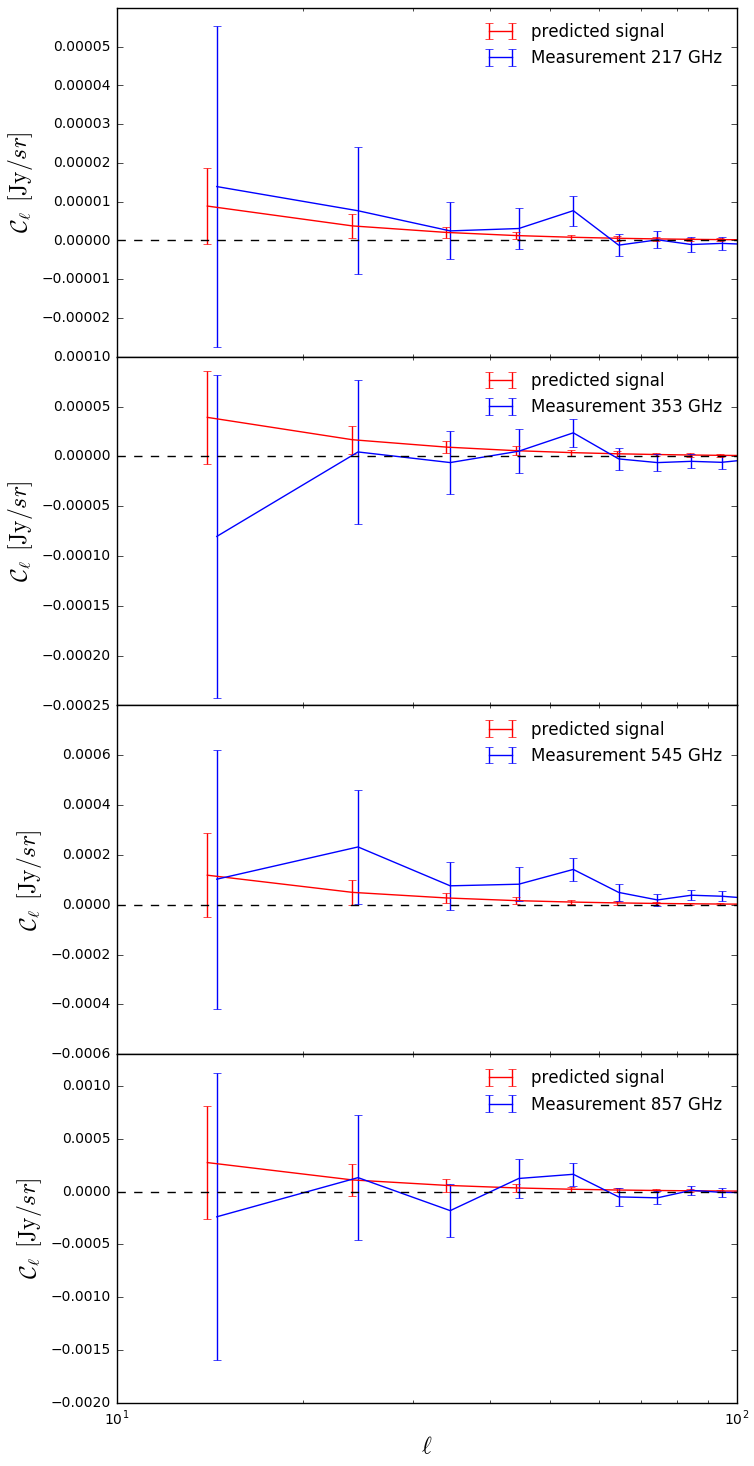}
\caption{Measured \mbox{CIBxCMB} cross-correlation and corresponding error bars for the GASS CIB field (blue points). The predicted signal (given by Eq.~(\ref{eq:clcr})) is shown in red. The error bars on the predicted signal are calculated using the noise term in Eq.~(\ref{eq:snr_real_bin}). Following \cite{Planck_cib_2014}, we assumed that the CIB power spectrum is contaminated by dust residuals at the level of 10\% when calculating the theoretical error bars.}
\label{fig:gass_545}
\end{figure}

Galactic dust emission is observed to be directly proportional to the HI gas column density in diffuse parts of the sky \citep[e.g.][]{boulanger_1996}. This empirical relation is used to clean the map from dust contamination in those parts of the sky \citep{Planck_2011}. We used the CIB maps derived in the GASS field  at the Planck HFI frequencies as our CIB input maps \citep{Planck_cib_2014}. GASS \citep{McClure-Griffiths_2009,Kalberla_2010} is a 21-cm line survey and is one of the most sensitive, highest angular resolution large-scale galactic HI emission survey ever made in the southern portion of the sky. The GASS CIB maps cover almost 11\% of the sky and is one of the cleanest (in terms of dust residuals) large patch of the CIB sky, as discussed in \cite{Planck_cib_2014}\footnote{We preferred using the CIB maps built as detailed in \cite{Planck_cib_2014} rather than the CIB maps provided by the Planck collaboration {\tt PLA} and constructed with the procedure detailed in \cite{Planck_gnilc_2016}. This is because the Planck CIB maps which use the GNILC technique to clean the dust, suffer from an excessive cleaning at the lower multipoles where we are interested in and this results in CIB power removal at these scales. See the Appendix for more details.}. As mentioned earlier, we use the {\tt HEALPix} package to analyse the maps. The angular resolution of the CIB maps is $16.2^\prime$ (set by the GASS survey angular resolution.)\footnote{The CIB maps, obtained with the PR1 Planck data release, have been corrected to the absolute calibration of the PR2 data release.}\\
We used the CMB map provided by {\tt PLA} generated using the SMICA component separation algorithm \citep{Cardoso_2008}. The CMB map is degraded to the same angular resolution of $16.2^\prime$ as that of the GASS CIB maps. We applied the GASS field mask ("Mask2" in Table~2 of \cite{Planck_cib_2014}) to both the CIB and CMB maps and cross-correlate them using {\tt Xpol} at all frequencies. Masking induces correlations between different multipoles and binning the power spectra helps to reduce these correlations. The measurements are then obtained as binned power spectra with a binning of $\Delta_\ell = 10$. \\
Figure E.6 of \cite{Planck_2016_cmbdiffuse} shows that the galactic dust and the CIB residuals in the SMICA CMB maps are quite low ($\le 0.002\%$). Hence, we don't expect a cross-correlation between the SMICA CMB map and the GASS CIB maps due to residual dust or residual CIB in the CMB maps.\\
Fig.~\ref{fig:gass_545} shows the observed cross-correlation compared to the theoretical prediction for the ISW signal at 217, 353, 545 and 857\,GHz. The theoretical prediction for ISW power spectrum is calculated using Eq.~(\ref{eq:clcr}). This power spectrum is then used to create a map which is given as an input to {\tt Xpol} to get the binned theoretical prediction for the ISW effect which is shown as the red curve in Fig.~\ref{fig:gass_545}. Just from the visual inspection it is hard to say if the observed signal is consistent with the predictions and if yes, then at what level. In order to understand the observed signal, we performed some tests that are discussed below.

\subsection{Significance with respect to the $\Lambda$CDM prediction} 

In order to calculate the significance of the measured cross-correlation $\hat{C}_\ell^{cr}$ with respect to the ISW $C_\ell^{cr}$ predicted by the $\Lambda$CDM model, we perform the following $\chi^2$ test. We estimate the amplitude $A$ such that $AC_\ell^{cr}$ is the best-fit solution to $\hat{C}_\ell^{cr}$, i.e. the value that minimize the following quantity:
\begin{equation}\label{eq:chi2}
\chi^2(A) = {\left[\hat{C}_\ell^{cr}(\nu) - AC_\ell^{cr}(\nu)\right]}^T \mathrm{Cov}_{cr}^{-1}(\nu)\left[\hat{C}_\ell^{cr}(\nu)AC_\ell^{cr}(\nu)\right]
\end{equation}
where $\mathrm{Cov}_{cr}(\nu)$ is the expected covariance matrix on the measured signal. The same minimization also gives us the error $\sigma_A$ on this amplitude. Such a test is convenient, as it informs us about: \textit{i)} whether the measured signal is compatible with our expectations, i.e. $A\sim1$; \textit{ii)} whether the measured signal is significant, i.e. $A/\sigma_A \gg 1$.

We used a Monte Carlo approach to compute the error bars for every bin and covariance between different bins (i.e. $\mathrm{Cov}_{cr}(\nu)$). This approach is similar to that described in, e.g., \cite{Giannantonio_2008,Serra_2014}. We simulated 1000 pairs of CIB and CMB maps which are correlated as expected from theory. The expected shot noise and galactic dust residual terms are added to the CIB maps. The detailed procedure is as follows:
\begin{itemize}
\item
A CIB map for a given frequency is created using the {\tt synfast} routine from {\tt HEALPix}. This map is the sum of the pure dust-free CIB term, shot noise and the dust residual term for the given frequency. Therefore, the total input power spectra to {\tt synfast} is:
\begin{equation}
C_\ell^\mathrm{CIB}(\nu) = C_\ell^\mathrm{CIB,pure}(\nu) + C_\ell^\mathrm{SN} + C_\ell^\mathrm{dust}(\nu)
\end{equation}
where 
$C_\ell^\mathrm{CIB,pure}(\nu)$, $C_\ell^\mathrm{SN}$ and $C_\ell^\mathrm{dust}(\nu)$ are the CIB, shot-noise and the dust-residual power spectra for a given frequency, respectively. The CIB power spectra are obtained from {\tt CAMB}. We assumed that after cleaning the GASS field using the dust to gas relation, 10\% dust residuals are left. Hence, we take the 10\% of the dust power spectra calculated as mentioned in Sect.~\ref{ssec:3.2}. The shot noise was derived from the \cite{Bethermin_2013} model. 
\item
The CMB map is created using the pure CMB term
\begin{equation}
C_\ell^\mathrm{CMB} = C_\ell^\mathrm{CMB,pure} 
\end{equation}
where the $C_\ell^\mathrm{CMB,pure}$ is obtained from {\tt CAMB}. 
\item
The CIB and CMB maps are created and correlated using {\tt synfast}. As explained before, the CIB map is created by adding the maps with the power spectra $C_\ell^\mathrm{CIB,pure}(\nu)$, $C_\ell^\mathrm{SN}$ and $C_\ell^\mathrm{dust}(\nu)$. Using the same seed used to create $C_\ell^\mathrm{CIB,pure}(\nu)$ map, we created an other map with the power spectra ${(C_\ell^\mathrm{cr})}^2/C_\ell^\mathrm{CIB,pure}$. Here, $C_\ell^{cr}$ is the predicted cross-correlation between the CIB and CMB through ISW effect given by Eq.~(\ref{eq:clcr}) and is again obtained from {\tt CAMB}. This second map is added to a third map created using a new seed and with power spectra $C_\ell^\mathrm{CMB} - {(C_\ell^\mathrm{cr})}^2/C_\ell^\mathrm{CIB,pure}$. Finally, these two maps will have amplitudes:
\begin{equation}
\begin{split}
&a_{\ell m}^\mathrm{CIB} = \xi_a{(C_\ell^\mathrm{CIB})}^{1/2}, \\
&a_{\ell m}^\mathrm{CMB} = \xi_a{\left({(C_\ell^\mathrm{cr})}^2/C_\ell^\mathrm{CIB,pure}\right)}^{1/2} + \xi_b{\left(C_\ell^\mathrm{CMB} - {(C_\ell^\mathrm{cr})}^2/C_\ell^\mathrm{CIB,pure}\right)}^{1/2}
\end{split}
\end{equation}
where $\xi$ is a random amplitude with zero mean and unit variance so that $\langle \xi \xi^*\rangle = 1$ and $\langle \xi\rangle = 0$.
This gives us:
\begin{equation}
\begin{split}
\left \langle a_{\ell m}^\mathrm{CIB}\,a_{\ell m}^\mathrm{CIB*} \right \rangle &= C_\ell^\mathrm{CIB}, \\
\left \langle a_{\ell m}^\mathrm{CIB}\,a_{\ell m}^\mathrm{CMB*} \right \rangle &= C_\ell^\mathrm{cr}, \\
\left \langle a_{\ell m}^\mathrm{CMB}\,a_{\ell m}^\mathrm{CMB*} \right \rangle &=  C_{\ell}^\mathrm{CMB} \,.
\end{split}
\end{equation}
\item
Once these maps are obtained, the GASS field mask covering $\sim$11\% sky is applied to both of them and their cross-power spectra are computed using {\tt Xpol}. This gives us 1000 sets of cross-spectra for each frequency which provide us with the uncertainty on our estimate. The covariance matrix for the binned power spectrum is then calculated as:
\begin{equation}
C_{bb'}^{cr} = {\left \langle (C_b - {\langle C_b\rangle}_\mathrm{MC})(C_{b'} - {\langle C_{b'}\rangle}_\mathrm{MC})\right \rangle}_\mathrm{MC}
\end{equation}
where $\langle . \rangle$ denotes the average of the Monte-Carlo samples. The error bars on every bin then becomes:
\begin{equation}
\sigma_{C_b} = {(C_{bb})}^{1/2}
\end{equation}
The error bars calculated this way can be compared with the analytical estimate:
\begin{equation}\label{eq:snr_real_bin}
\sigma_{C_b}(\nu) = {\left(\dfrac{C_b^{cr}(\nu)^2 + \left(C_b^{\mathrm{CIB}}(\nu) + N_b^{\mathrm{CIB}}(\nu)\right)\, C_b^{\mathrm{CMB}}}{f_\mathrm{sky}(2\ell + 1) \Delta\ell}\right)}^{1/2}
\end{equation}
We observe that the error bars obtained from the simulations are 22-25\% higher than those obtained with Eq.~(\ref{eq:snr_real_bin}) for different frequencies. This is expected as Eq.~(\ref{eq:snr_real_bin}) is not very accurate for very low sky fractions which is the case here. We calculate the correlation matrix based on the 1000 simulations for every frequency and find that the correlations between adjacent multipoles are always less than 16\% (with a binning $\Delta_\ell = 10$). 
\end{itemize}
Once we have obtained the expected covariance matrix using the above procedure, we can use Eq.~(\ref{eq:chi2}) to minimise the $\chi^2$ and fit for the amplitude $A$. The results are shown in Table~\ref{tab:chi2}. Apart from 545\,GHz, we see that although the amplitudes of the measured signals are roughly compatible with 1 (within the estimated error bars), the significance of those cross-correlations is quite low.  At 545\,GHz, $A/\sigma_A$ is equal to 5.52 with $A = 6.64$. Therefore, the observed amplitude of the cross-correlation is $(6.64 -1)/1.2 = 4.7\sigma$ higher than what is expected from the $\Lambda$CDM model. Possible explanations for this high significance are explored in the next subsection.

\begin{table}[t]
\centering
\begin{tabular}{c|ccc|ccc}
\hline
\hline 
& \multicolumn{3}{c|}{with $\ell = 55$} &  \multicolumn{3}{c}{without $\ell = 55$} \\
\hline
\begin{tabular}{@{}c@{}}Freq.\\ (GHz)\end{tabular} &  $A$ & $\sigma_A$ & $A/\sigma_A$ &  $A$ & $\sigma_A$ & $A/\sigma_A$\\ 
\hline
 217 & 2.03 & 1.12 & 1.8 & 0.77 & 1.24 & 0.63 \\
 353 & -0.19 & 1.12 & -0.17 & -1.22 & 1.22 & -1.0 \\
 545 & 6.64 & 1.20 & 5.52 & 5.53 & 1.34 & 4.12 \\
 857 & 0.11 & 1.62 & 0.07 & -1.17 & 1.80 & -0.65 \\
\hline
\end{tabular}
\newline
\centering \caption{Best-fit values and errors for the amplitude $A$, and significance of the signal $A/\sigma_A$ for all the frequencies. These have been calculated using $\ell \in$ [10, 100], with and without the data point at $\ell = 55$ (see text for more details). No significant detection of the signal with respect to the $\Lambda$CDM prediction can be claimed based on these values apart from the signal observed at 545\,GHz.}\label{tab:chi2}
\end{table}


\subsection{Null test}

\begin{figure}[h]
\centering
\includegraphics[width=9cm]{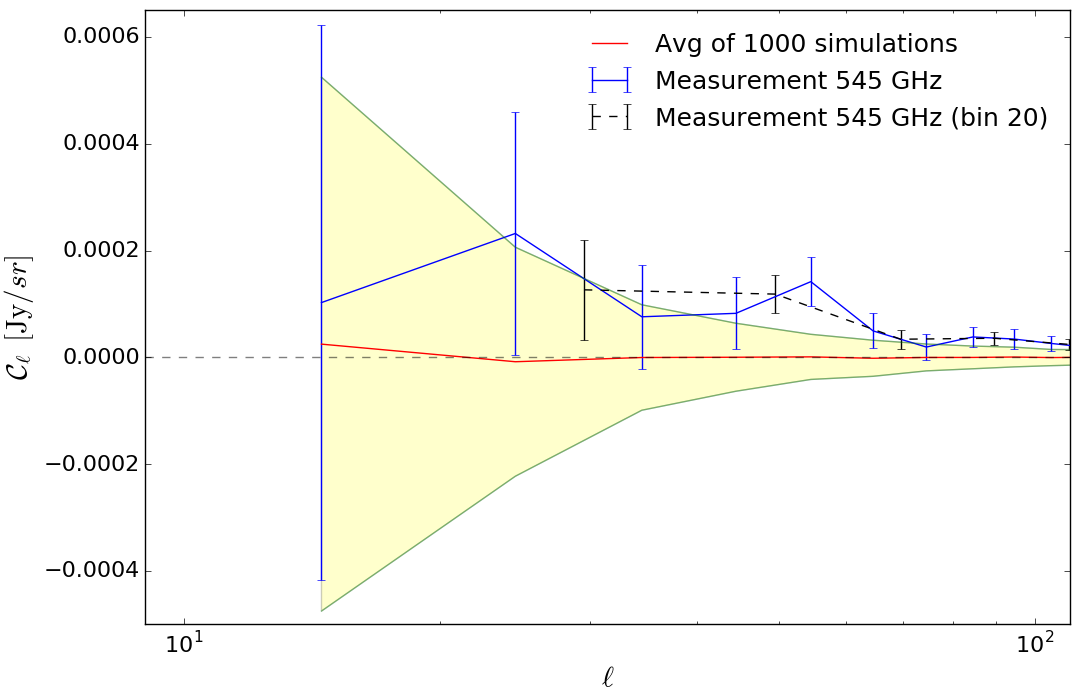}
\caption{The average null signal calculated using the cross-correlation of the GASS CIB field with 1000 random CMB maps is shown in red. The shaded area in yellow spans the 1$\sigma$ region of the null signal which is represented by the green lines. The measured \mbox{CIBxCMB} cross-correlation is shown in blue. We also show the \mbox{CIBxCMB} cross-correlation measured using a bin size $\Delta\ell$=20 instead of 10 (black points and dash line).}
\label{fig:nulltest}
\end{figure}

The previous test showed that no conclusion can be reached regarding the measured cross-correlations being consistent with a detection or not. Therefore, we also checked the significance of the observed signal against the null hypothesis. For this purpose, we performed a null test: for every frequency, we cross-correlated the GASS CIB field map with 1000 randomly generated CMB maps. These 1000 cross-correlations give us the covariance matrix and hence also the error bars for different bins for the null signal. Fig.~\ref{fig:nulltest} shows the 1$\sigma$ region spanned by this null signal, together with the measured \mbox{CIBxCMB} cross-correlation at 545\,GHz.
 
We perform a $\chi^2$ test for the observed cross-correlation signal with the null signal such that 
\begin{equation}
\chi^2_\mathrm{null} = {\left[\hat{C}_\ell^{cr}(\nu)\right]}^T \mathrm{Cov}_{cr}^{-1}(\nu)\left[\hat{C}_\ell^{cr}(\nu)\right]
\end{equation} 
where $\hat{C}_\ell^{cr}(\nu)$ is the observed cross-correlation signal and $\mathrm{Cov}_{cr}(\nu)$ is the covariance matrix computed from the simulations as mentioned above for all the frequencies. 
We get a $\chi^2$ value of 6.0, 5.6 and 4.9 for 9 degrees of freedom which corresponds to the p-value of 0.74, 0.78 and 0.84 for 217, 353 and 857\,GHz respectively which shows that the signal is compatible with the null. For 545\,GHz, we get a $\chi^2$ of 30.7 which corresponds to a p-value of 0.0003. Such high value of the $\chi^2$ is observed because of a "bump" observed in the measured cross correlation near $\ell \approx 55$. $\chi^2$ drops down to 16.7 with a corresponding p-value of 0.03 when this data point is not considered while calculating the $\chi^2$. \\

We investigated the source of the "bump" seen at  $\ell \approx 55$. It can have three different origins: residual systematics in the data, correlated residual astrophysical signals (e.g. galactic dust),  or a simple $<3\sigma$ deviation. The bump is observed at all frequencies (from 217 to 857\,GHz), which tends to rule out any systematics left in the data. In the GASS field, it is also present when using the GNILC CIB map (see Appendix) rather than the Planck 2013 CIB map, and also using different Planck CMB maps (SMICA but also COMMANDER).  On the contrary, it is not visible in an other field when computing the cross-correlation using the GNILC CIB map and 11\% of the sky in the northern hemisphere. This tends to show that it cannot be due to a residual astrophysical signal correlated in the CIB and CMB maps, and is thus a feature, which is not statistically significant (less than 3$\sigma$), in the GASS field.  \\

From the above tests, we see that the cross-correlation between the CIB and CMB obtained for the GASS field (with a sky fraction of 11\%) is compatible with the null signal. We need CIB maps extracted over larger sky fractions and with lower dust residuals to extract the ISW signal using the \mbox{CIBxCMB} cross-correlation.


\section{Conclusions}\label{sec:6}

We used the CIB as a potential tracer of the large scale structures to extract the ISW signal present in the CMB with a high signal-to-noise ratio. Our approach makes use of the framework presented in \cite{Ilic_2011}. We used the improved linear CIB model of \cite{Maniyar_2018} which includes the \mbox{CIBxCMB} lensing cross-correlation while fitting for the data which partially breaks the degeneracy between the effective bias of the CIB galaxies and their emissivity and provides improved constraints on the CIB model parameters.

With this improved CIB model, we computed theoretical predictions for the \mbox{CIBxISW} cross-correlation for different frequencies and find -- as expected from the ISW effect -- that it peaks at large angular scales, namely 20 $< \ell <$ 50 for all the frequencies considered. We showed that this cross-correlation can provide us the ISW signal with the highest SNR ever obtained for a single tracer. The optimal frequency with the highest SNR for the ISW detection in this case is 857\,GHz with an SNR of 6.7$\sigma$. Using a Fisher matrix approach, we calculated the expected improvement on the constraints on the cosmological parameters as set by Planck ($+$SN, BAO, and H0) by adding the ISW signal obtained with the \mbox{CIBxCMB} cross-correlation. This analysis was done for the $\Lambda$CDM and the $w$CDM cosmological models, and we found as expected that the constraints on the later are more sensitive to the addition of the ISW signal. For that model, we found that the \mbox{CIBxCMB} cross-correlation could potentially improve the constraints on the dark energy density parameter $\Omega_\Lambda$ by $\sim30$\% and the equation of state of the dark energy $w$ by $\sim80$\%, whether the available fraction of the sky is 70\% or 40\%. It also improves the constraints on the $A_s$ parameter, as well as $\tau$ due to their degeneracy.  The \mbox{CIBxCMB} lensing cross-correlation breaks the degeneracy between $A_s$ and $\tau$ and improves the constraints on these parameters. The so-called ``Stage-4'' next-generation ground-based CMB experiment (CMB-S4) is expected to produce maps of unprecedented quality, improving the signal-to-noise ratio of the reconstructed lensing map by over an order of magnitude \citep{CMBS4_2016} compared to Planck. While this will naturally improve the signal-to-noise ratio for the \mbox{CIBxCMB} lensing cross-correlation, one should keep in mind that the maximum amount of cosmological information that can be extracted is ultimately limited by cosmic variance. We performed a Fisher matrix analysis to forecast the constraints on the cosmological parameters with the improved \mbox{CIBxCMB} lensing cross-correlation. As expected, compared to current constraints using Planck data, a small improvement (a few \%) is observed for most cosmological parameters and a significant improvement ($\sim$ 30\%) for $\tau$ and A$_s$.

This analysis was however performed in relatively idealized scenarios with CIB maps free of any galactic dust residuals. To assess the effects of partial sky coverage and galactic dust residuals, we calculated the dust power-spectra for all the frequencies over various sky fractions. We find that at our multipoles of interests in i.e. $\ell < 100$, the residual dust power spectra are completely dominating over the CIB power spectra. As a result, taking 10\% of the dust power spectra as the possible residuals in our cleaned maps, we performed the same analysis over given portions of the sky and show that the SNR comes down drastically with e.g. a SNR of 1.5 for 353\,GHz over 20\% sky fraction.

Using CIB and CMB maps obtained by Planck we attempted to measure the \mbox{CIBxCMB} cross-correlation, allowing us to check the feasibility of detecting the ISW effect with the current level of dust removal and available sky fraction. This cross-correlation was computed over $\sim$11\% of the southern sky in a field which is one of the largest and cleanest patch of the CIB to date \citep{Planck_cib_2014}. We calculated the significance of the observed signal with respect to the $\Lambda$CDM prediction. We performed Monte-Carlo simulations in order to obtain the expected covariance matrix between different bins for a given frequency for this test. We find that for all the frequencies apart from 545\,GHz, the signal is roughly compatible with expectation, but is detected with very low significance. At 545\,GHz, the signal is detected at 5.5$\sigma$ but at the same time, its amplitude is 4.7$\sigma$ higher than that predicted by the $\Lambda$CDM model. We also performed a null-test to see if the detected signal is compatible with the null hypothesis. Again, apart from 545\,GHz, we find that the signal is consistent with the null hypothesis. For 545\,GHz, the null hypothesis is rejected with the p-value of 0.0003, with a $\chi^2$ of 30.7 for 9 degrees of freedom. Performing the null test by ignoring a bump observed in the cross-correlation around $\ell \sim 55$ gives a p-value of 0.03 with a $\chi^2$ of 16.7 for 8 degrees of freedom. These tests show that the measured \mbox{CIBxCMB} cross-correlation is compatible with the null hypothesis and that we do not detect any significant signal. Our analysis shows that with the current state of the maps, the ISW signal cannot be detected. A much better cleaning of galactic dust is required. If we do not want to degrade the SNR on the measurements by more than 10\% on 40\% of the sky, we need to clean the dust up to the 0.01\% level on the power spectrum. This is very challenging (if at all possible) to achieve but mandatory for using the CIB as a cosmological probe -- either through the ISW effect but also in itself, e.g. for constraining local primordial non-Gaussianity or other non-standard cosmological models. \cite{Granett_2008} performed a stacking of CMB patches at the locations of a sample of 100 large voids and galaxy clusters, aiming at measuring their ISW signal. They reported a signal whose amplitude was significantly larger than expected in the $\Lambda$CDM model. This work has later on been challenged by other groups, and extended to larger catalogues of structures \citep[see e.g.][and references therein for a recent review]{Kovacs_2018}, with a consensus that the stacking signal appears indeed larger than expected. If the origin of this discrepancy were to be an excess of ISW signal, it is not straightforward to predict how (or if) it would appear in our present analysis, due to us working in harmonic space whereas stacking is done directly in maps. We intend to explore this particular subject in the near future in order to determine if this potential excess ISW signal could be detected using the CIB maps along with the CMB.

\begin{acknowledgements}
A.M. and G.L. warmly thank Matthieu Tristram for his invaluable help in the use of Xpol. We acknowledge financial support from the \textquotedblleft Programme National de Cosmologie and Galaxies" (PNCG) funded by CNRS/INSU-IN2P3-INP, CEA and CNES, France, from the ANR under the contract ANR-15-CE31-0017 and from the OCEVU Labex (ANR-11-LABX-0060) and the A*MIDEX project (ANR-11-IDEX-0001-02) funded by the \textquotedblleft Investissements d'Avenir" French government programme managed by the ANR. The research leading to these results has also received funding from the European Research Council under the European Union's Seventh Framework Programme (FP7/2007-2013)/ERC Grant Agreement No. 617656 ``Theories and Models of the Dark Sector: Dark Matter, Dark Energy and Gravity."
\end{acknowledgements}

\bibliographystyle{aa}
\bibliography{bib2}


\appendix\label{app}

\section{Choice of the CIB maps}

As mentioned in Sect.~\ref{ssec:5.1}, we used the CIB maps in the GASS field built as detailed in \cite{Planck_cib_2014}, and not the GNILC CIB maps provided by the Planck collaboration. The Planck CIB maps have been prepared by cleaning the galactic dust using the GNILC component separation method and the details are provided in \cite{Planck_gnilc_2016}. Fig. 6 of \cite{Planck_gnilc_2016} shows the power spectrum of galactic dust and galactic-dust subtracted Planck maps (a.k.a the CIB), together with the best-fit model of the CIB power spectrum from \cite{Planck_cib_2014}, for different frequencies. This figure shows that there is an excessive dust cleaning at 545 and 857\,GHz at low multipoles which are of interest to us and hence the CIB power spectra from GNILC maps is lying below \cite{Planck_cib_2014} best-fit model. \cite{Maniyar_2018} provided an improved CIB model, using not only the CIB measurements but also the \mbox{CIBxCMB} lensing cross-correlation. \\

To check the validity of the CIB map in the GASS field, we cross-correlate it with the lensing convergence $\kappa$ map (provided by Planck\footnote{\url{https://wiki.cosmos.esa.int/planckpla2015/index.php/Specially_processed_maps\#2015_Lensing_map}}) using {\tt Xpol}. For comparison, we also considered the Planck GNILC CIB map from PLA and cross-correlate it with the same lensing convergence map. The cross-correlation of \mbox{CIBxCMB} lensing potential $\phi$ is obtained as:
\begin{equation}
C_\ell^{\mathrm{CIB} \times \phi} = \dfrac{C_\ell^{\mathrm{CIB} \times \kappa}} {0.5\ell (\ell +1)}
\end{equation}

\begin{figure}[h]
\centering
\includegraphics[width=9cm]{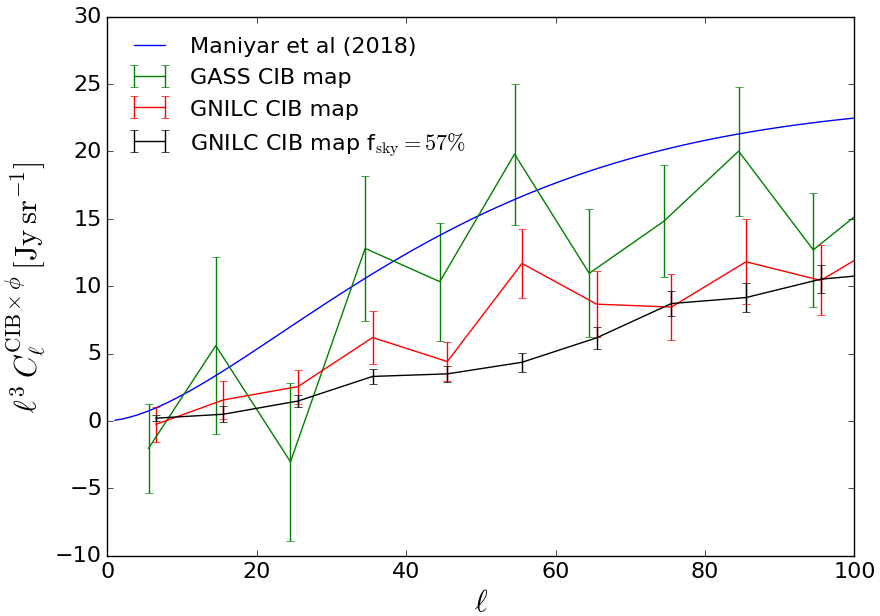}
\caption{\mbox{CIBxCMB} lensing cross power spectra for $2\le \ell \le 100$ at 545\,GHz. The blue curve shows the \mbox{CIBxCMB} lensing model from \cite{Maniyar_2018}. The green, red and black curves show the measured  cross-correlation for GASS CIB map, GASS GNILC map, and GNILC maps on 57\% of the sky, respectively. It is observed that the cross-correlation is consistently below the \cite{Maniyar_2018} model for the GNILC CIB map. The cross-correlation obtained with the GASS CIB map is in better agreement with the \cite{Maniyar_2018} model.}
\label{fig:cib-phi}
\end{figure}

Figure\,\ref{fig:cib-phi} shows the $\ell^3\:C_\ell^{\mathrm{CIB} \times \phi}$ computed both for GASS CIB map and Planck GNILC CIB map for 545\,GHz till $\ell = 100$. We see that the cross-correlation using the GASS CIB map falls on \cite{Maniyar_2018} model whereas the cross-correlation with the Planck GNILC CIB map is consistently lower (by a factor of $>2$ at $\ell$=50) in this range of multipoles where we are interested in for ISW. At the same time, it should be noted that the error bars on the cross-correlation with the Planck GNILC CIB map are smaller than the error bars on the cross-correlation with the GASS CIB map. From Eq.~(\ref{eq:snr_real_bin}), we can infer that this is because the amount of the dust in the Planck GNILC CIB maps is lower than the GASS CIB maps. These conclusions hold at 857 and 353\,GHz, where the measurements are more than 2 times lower than what is expected from the model at $\ell$=50.\\

\begin{figure}[h]
\centering
\includegraphics[width=9cm]{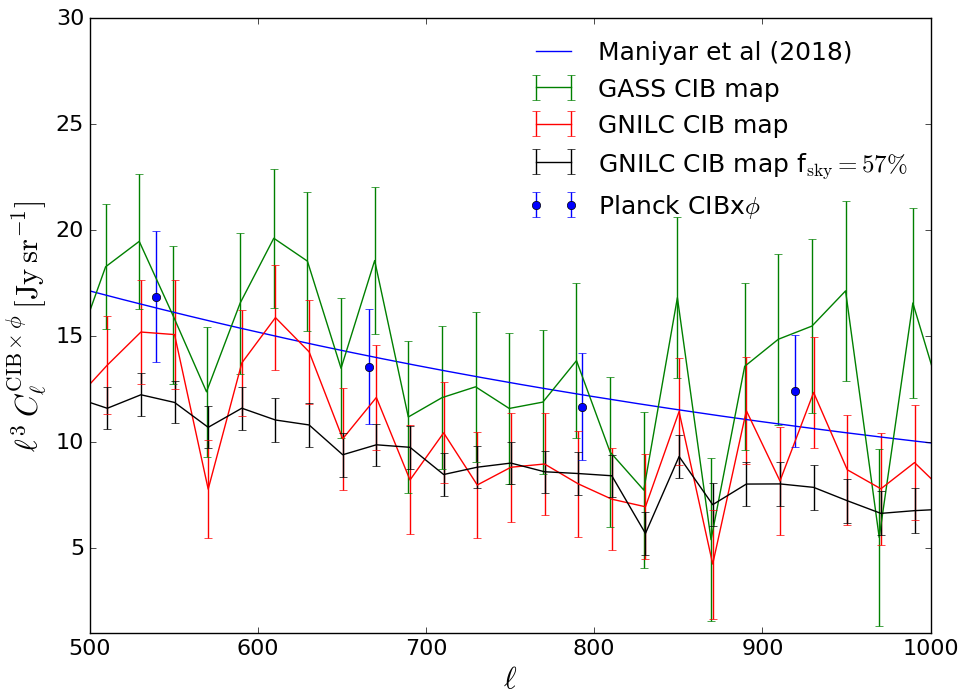}
\caption{\mbox{CIBxCMB} lensing cross power spectra for $500\le \ell \le 1000$ at 545\,GHz. The blue curve shows the \mbox{CIBxCMB} lensing cross-correlation model by \cite{Maniyar_2018}. The green and red curves show the measured cross-correlation in the GASS field for the CIB maps from \cite{Planck_cib_2014} and GNILC, respectively. The black curve is the cross-correlation using GNILC CIB map on 57\% of the sky. Data points in blue show the CIB$\times \phi$ measurements by Planck \citep{Planck_ciblensing_2014}. It is observed that the cross-correlation for the GASS CIB map is in better agreement with \cite{Maniyar_2018} model than for the Planck GNILC CIB map. }
\label{fig:cib-phi-2}
\end{figure}

Middle panel of the Fig. 6 of \cite{Planck_gnilc_2016} shows that for 545\,GHz, the best fit Planck CIB and the GNILC CIB power spectra start agreeing only after $\ell \sim 800$. Figure.\,\ref{fig:cib-phi-2} shows the $\ell^3\:C_\ell^{\mathrm{CIB} \times \phi}$ computed both for GASS CIB map and Planck GNILC CIB map for 545\,GHz from $\ell = 500$ to $\ell = 1000$. It can be observed that overall, although the cross-correlation with the GASS CIB map is in better agreement with \cite{Maniyar_2018} model than with the Planck GNILC CIB map, GNILC CIB maps perform better at these multipoles than at multipoles $< 100$ (but still systematically underestimate the cross-correlation). This once again shows that there is an excessive cleaning of galactic dust in the Planck GNILC CIB maps, which results in partial CIB removal and hence an underestimate of the the CIB power. That is the reason why we decided not to use this CIB map in our analysis. 

\end{document}